\documentclass[aps, prd, onecolumn, groupedaddress, nofootinbib, preprintnumbers, 11pt]{revtex4-2}

\usepackage{exdm_user_manual}

\begin{document}

\title{EXCEED-DM: Extended Calculation of Electronic Excitations for Direct Detection of Dark Matter}

\author{Tanner Trickle}
\email{ttrickle@fnal.gov}
\affiliation{Theory Division, Fermi National Accelerator Laboratory, Batavia, Illinois, 60510, USA}
\preprint{FERMILAB-PUB-22-767-T}

\date{\today}

\begin{abstract}
    Direct detection experiments utilizing electronic excitations are spearheading the search for light, sub-GeV, dark matter (DM). It is thus crucial to have accurate predictions for any DM-electron interaction rate in this regime. \textsf{EXCEED-DM}~\GithubLink (EXtended Calculation of Electronic Excitations for Direct detection of Dark Matter) computes DM-electron interaction rates with inputs from a variety of \textit{ab initio} electronic structure calculations. The purpose of this manuscript is two-fold: to familiarize the user with the formalism and inputs of \textsf{EXCEED-DM}, and perform novel calculations to showcase what \textsf{EXCEED-DM} is capable of. We perform four calculations which extend previous results: the scattering rate in the dark photon model, screened with the numerically computed dielectric function, the scattering rate with an interaction potential dependent on the electron velocity, an extended absorption calculation for scalar, pseudoscalar, and vector DM, and the annual modulation of the scattering rate in the dark photon model.
\end{abstract}

\maketitle
\newpage
\tableofcontents
\newpage

\section{Introduction}
\label{sec:introduction}

Discovering the nature of dark matter (DM) is one of the most important goals in all of physics. Constituting $26.5 \%$ of the universe's energy density~\cite{ParticleDataGroup:2020ssz}, DM is roughly five times more abundant than ordinary matter; yet any understanding of its particle content eludes us. One class of ongoing experiments, which search for DM-Standard Model interactions in a laboratory, are known as direct detection experiments. The canonical example of a direct detection experiment looks for nuclear recoils induced by a scattered DM particle. Many of these experiments have been performed, or are proposed, e.g., LUX~\cite{LZ:2015kxe}, SuperCDMS~\cite{SuperCDMS:2014cds}, ANAIS~\cite{Amare:2019jul}, CRESST~\cite{CRESST:2020ghz}, SABRE~\cite{SABRE:2018lfp}, DAMA~\cite{Bernabei:2020mon}, PandaX~\cite{PandaX-II:2017hlx}, DAMIC~\cite{DAMIC:2016lrs}, NEWS-G~\cite{NEWS-G:2017pxg}, LZ~\cite{LZ:2015kxe}, XENON10/100/1T~\cite{XENON100:2012itz, XENON:2018voc, XENON:2007uwm}, KIMS~\cite{Kim:2015prm}, DM-Ice~\cite{Jo:2016qql}, DarkSide~\cite{DarkSide:2018bpj}, and the absence of any signal has led to stringent bounds on a wide variety of DM models. However, all of these experiments share a common kinematics problem when looking for DM with mass lighter than a GeV; when the DM mass drops below the target nucleus mass, the energy deposited in a scattering event rapidly falls below the detection threshold. 

This is an unfortunate problem because many well-motivated DM models can possess a light, sub-GeV, DM candidate (see Refs.~\cite{Knapen:2017xzo,Cooley2022,Essig2022} for recent reviews). This problem has been remedied by an extraordinary influx of new detection ideas for covering this region of DM parameter space (see Refs.~\cite{Essig2022,Mitridate2022,Essig2022a,Kahn:2021ttr} for recent reviews). Currently, the preeminent method for searching for sub-GeV DM is via electronic excitations across band gaps in crystal targets~\cite{Trickle:2019nya,Chen2022,Maity2020,Radick:2020qip,Tan:2021nif,Derevianko:2010kz,Dzuba:2010cw,Blanco:2021hlm,Heikinheimo:2019lwg,Emken:2019tni,Cavoto:2017otc,Mitridate:2021ctr,Gelmini:2020xir,Liang:2018bdb,Kadribasic:2017obi,Catena:2021qsr, Catena2022, Knapen:2021run,Blanco:2019lrf,Knapen:2021bwg,Hochberg:2021pkt,Griffin:2021znd,Hochberg:2016ntt,Hochberg:2017wce,Bloch:2016sjj,Derenzo:2016fse,Essig:2011nj,Essig:2015cda,Graham:2012su,Griffin:2019mvc,Griffin:2020lgd,Hochberg:2015fth,Hochberg:2015pha,Coskuner:2019odd,Geilhufe:2019ndy,Hochberg:2016ajh,Hochberg:2016sqx,Kurinsky:2019pgb,Lee:2015qva}. With band gaps of $\mathcal{O}(\text{eV})$, DM with masses as light as $\mathcal{O}(\text{MeV})$ can be searched for via a scattering event, and $\mathcal{O}(\text{eV})$ masses can be searched for via absorption. Experiments such as SuperCDMS~\cite{SuperCDMS:2020ymb,SuperCDMS:2018mne,CDMS:2009fba}, DAMIC~\cite{DAMIC:2020cut,DAMIC:2016qck,Settimo:2020cbq,DAMIC:2015znm,DAMIC:2019dcn}, EDELWEISS~\cite{EDELWEISS:2018tde,EDELWEISS:2019vjv,EDELWEISS:2020fxc}, SENSEI~\cite{Crisler:2018gci,SENSEI:2020dpa,SENSEI:2019ibb}, and CDEX~\cite{Zhang2022} are currently in operation searching for DM induced electronic excitations with a combination of Silicon (Si) and Germanium (Ge) targets. Similar targets utilizing electronic excitations with smaller, $\mathcal{O}(\text{meV})$, band gaps such as spin-orbit coupled materials~\cite{Inzani:2020szg,Chen2022}, Dirac materials~\cite{Hochberg:2017wce,Coskuner:2019odd,Geilhufe:2019ndy}, and superconductors~\cite{Hochberg:2015fth,Hochberg:2015pha,Hochberg:2016ajh} have also been proposed. 

The calculation of DM-electron interaction rates in these targets is more involved than the standard nuclear recoil calculation. This is because electrons in crystal targets cannot be treated as free states. The crystal creates a lattice potential, distorting the electronic wave functions. Therefore, estimates of DM-electron interaction rates are crucially dependent on how the electronic wave functions, and band structure, are modeled. With experiments already being performed, and the nature of DM still unknown, it is crucial to have the tools to make predictions for the broadest set of DM models~\cite{Kahn2022a, Catena2022}. The earliest calculations used analytic approximations of the wave functions, along with the measured density of states to account for the band structure~\cite{Graham:2012su, Lee:2015qva}. Then the \textsf{QEDark}~\cite{Essig:2015cda} program was introduced, and was the first to incorporate electronic wave functions, and band structures, computed with density functional theory (DFT) in calculations of the DM-electron scattering rate. This was a monumental step forward, creating a connection between first principles condensed matter calculations and particle physics observables which exists today~\cite{Mitridate2022, Kahn:2021ttr}. Additionally, an extension of \textsf{QEDark}, \textsf{QEDark-EFT}~\cite{Catena:2021qsr,Catena2022}, has been introduced to generalize the set of DM models for which DM-electron scattering can be computed. Recently it was shown that, in some simple DM models, the DM-electron interaction rate could be related to the ($\mathbf{q}, \omega$ dependent) dielectric function~\cite{Knapen:2021run,Hochberg:2021pkt}. This rewriting rigorously included screening effects, which had previously been assumed to be small. While projections for the DM-electron interaction rate in these models were made with DFT calculations and simplified models of the dielectric function, the longer term vision is to use the experimentally measured dielectric function to circumvent the uncertainties associated with DFT calculations. The program~\textsf{DarkELF}~\cite{Knapen:2021bwg} was introduced to compute DM-electron interaction rates from input dielectric functions.

In parallel with these developments, \textsf{EXCEED-DM}~\GithubLink (\textbf{EX}tended \textbf{C}alculation of \textbf{E}lectronic \textbf{E}xcitations for \textbf{D}irect detection of \textbf{D}ark \textbf{M}atter)~\cite{Griffin:2021znd, Trickle2022} has also been developed. Similar to \textsf{QEDark}, \textsf{EXCEED-DM} can utilize \textit{ab initio} DFT calculations of the target electronic structure. Previous versions of \textsf{EXCEED-DM} have been used to compute DM-electron scattering in a variety of target materials~\cite{Trickle:2019nya, Griffin:2021znd, Trickle:2019nya}, DM-electron absorption for a variety of DM models~\cite{Mitridate:2021ctr}, as well as DM-electron interaction rates in more novel, spin-orbit coupled targets~\cite{Chen2022}. With similar goals as the previously discussed programs, it is important to distinguish how \textsf{EXCEED-DM}, specifically the newest version, \textsf{v1.0.0}, differentiates itself. \textsf{EXCEED-DM} has three main advantages relative to \textsf{QEDark}, the current standard for these calculations:
\begin{itemize}
    \item \textbf{Independent of Electronic Structure Calculator}. \textsf{QEDark} is intrinsically linked to \textsf{Quantum Espresso}~\cite{Giannozzi2009,Giannozzi2017}, a specific program for performing DFT calculations of the electronic structure, i.e., the electronic wave functions and band structure. Therefore, the accuracy of the electronic structure is limited to the methods used within \textsf{Quantum Espresso}. \textsf{EXCEED-DM} is not linked to a specific electronic structure calculator; if the output wave functions are in a format supported by \textsf{EXCEED-DM} (discussed more in the next bullet point, Sec.~\ref{subsec:elec_state_config}, and in the documentation~\DocLink) any electronic structure calculator can be used. For example, \textsf{EXCEED-DM} has been used with electronic wave functions in the plane wave (PW) basis computed with \textsf{Quantum Espresso} and \textsf{VASP}~\cite{Kresse1996,Kresse1999,Kresse1993,Kresse1994}. 
    \item \textbf{Variety of electronic state approximations}. \textsf{QEDark} performs calculations with the initial and final electronic states expanded in a PW basis. This basis works well for states close to the Fermi surface. For states farther away from the Fermi surface this approach is suboptimal since larger momentum cutoffs are needed, especially for deeply bound, core, states~\cite{Griffin:2021znd}. Therefore it is beneficial to represent electronic states in a variety of bases. As of \textsf{v1.0.0}, \textsf{EXCEED-DM} supports three different bases, the PW basis, a Slater type orbital (STO) basis, and a single PW, see Sec.~\ref{subsec:elec_state_config} for more detail. Any of these bases can be used to approximate the initial or final electronic states; moreover a combination of bases can be used within the set of initial and final states. These bases were chosen since they were previously used to extend the description of the electronic states in Si and Ge targets, see Ref.~\cite{Griffin:2021znd} for more details. Any electronic structure calculator, or combination of them, which can output wave functions in these bases can be used as input to \textsf{EXCEED-DM}. While \textsf{v1.0.0} only includes a few bases, these are meant to serve as examples for future versions, which will expand the number of bases included.\footnote{Until the basis sets are included in a versioned release it is up to the user to implement these in \textsf{EXCEED-DM}. However, an experienced \textsf{Fortran} programmer should be able to add new bases with relative ease following the currently implementations as examples.}
    \item \textbf{Additional calculations}. In addition to DM-electron scattering rate calculations, \textsf{EXCEED-DM} can perform DM-electron absorption calculations. While for some DM models, absorption on electrons can be related to the long wavelength dielectric function, $\varepsilon(0, \omega)$, for other models this is not possible~\cite{Mitridate:2021ctr, Chen2022}. Moreover, within DM-electron scattering rate calculations, \textsf{EXCEED-DM} can straightforwardly compute the daily modulation rate~\cite{Trickle:2019nya}; an important feature for discriminating against potential backgrounds in anisotropic targets.
\end{itemize}
Additionally, there are miscellaneous technical advantages, e.g., \textsf{EXCEED-DM} is parallelized with \textsf{OpenMPI}, versus \textsf{OpenMP} used in \textsf{QEDark}. This allows for calculations to utilize the many memory independent cores available in, for example, supercomputers, thereby increasing parallelization. Given enough cores, this will lead to faster DM-electron interaction rate calculations. In principle, all previous \textsf{QEDark} calculations can be performed with \textsf{EXCEED-DM} with appropriate conversion of \textsf{Quantum Espresso} output to \textsf{EXCEED-DM} input. This may be useful when comparing the results of previous calculations. 

The \textsf{QEDark-EFT} extension to \textsf{QEDark} can compute DM-electron scattering rates for a wider variety of DM models, under the same approximations about the electronic states. Therefore the main advantages of \textsf{EXCEED-DM} are still present relative to \textsf{QEDark-EFT}. While \textsf{EXCEED-DM} does not currently have routines for calculating DM-electron scattering rates for general operators, the primitive building blocks, discussed in Sec.~\ref{subsec:transition_matrix_elements}, are available. Moreover, \textsf{EXCEED-DM} is structured such that calculations of more general DM-electron scattering rates need only be written, and coded, in terms of transition matrix elements, $\mathcal{T} \equiv \langle F | \mathcal{O} | I \rangle$, in order to be added. This separates the pieces of the calculation dependent on the electronic state approximations, from the observables being computed. Adding new models is straightforward, and in Sec.~\ref{subsec:dme_VA} we discuss computing the DM-electron scattering rate via a scattering potential dependent on the electron velocity which arises from a very simple DM model.

\textsf{EXCEED-DM} serves a different purpose than \textsf{DarkELF}, which can compute DM-electron interaction rates, for some DM models, starting from the dielectric function. \textsf{EXCEED-DM} is compatible with \textsf{DarkELF} since it can compute the ($\mathbf{q}, \omega$ dependent) dielectric function, which can then be used as input to \textsf{DarkELF}. Additionally, \textsf{EXCEED-DM} can account for screening effects, the importance of which was emphasized in Refs.~\cite{Knapen:2021run,Hochberg:2021pkt}. There are two ways to account for screening in \textsf{EXCEED-DM}, via an analytic model of the dielectric function, or using a pre-computed dielectric function. An analytic model for screening was used previously in Ref.~\cite{Griffin:2021znd} and in Sec.~\ref{subsec:dme_SI_numeric_dielectric} we use the dielectric function computed with \textsf{EXCEED-DM} to screen the DM-electron interactions. However, this is not the only pre-computed dielectric function one can use; if available, a file containing the measured dielectric function can also be used to screen the interaction.

The goal of this paper is to familiarize a user with the types of calculations that can be performed with \textsf{EXCEED-DM}, first by introducing a formalism which encompasses all the calculations, and then detailing the inputs to \textsf{EXCEED-DM} and performing new calculations. While these new calculations will serve as examples, they are also new physics results in their own right, and are the most precise determination of DM-electron interaction rates in Si and Ge detectors to date. Our goal is not to derive all the relevant DM-electron interaction formulas from first principles (these are studied in detail in numerous other, referenced, works), but rather illustrate how \textsf{EXCEED-DM} can be used once these formulas are derived.

This paper is organized as follows, in Sec.~\ref{sec:formalism} we discuss the framework under which the \textsf{EXCEED-DM} calculations are performed. While the formalism for DM-electron interaction rates has been developed previously, it is often developed for different purposes, and therefore written in slightly different languages. Since \textsf{EXCEED-DM} can perform many different calculations it is important to have a common theoretical formalism for all the calculations. Moreover, this makes it clear the language in which future extensions, e.g., calculations for other DM models or different electronic state approximations, need to be written in for easy addition to \textsf{EXCEED-DM}. In Sec.~\ref{subsec:elec_state_config} we discuss the electronic configuration and the approximations regarding the initial and final states that can be used. In Sec.~\ref{subsec:transition_matrix_elements} we discuss the \textit{transition matrix element}, the key link between different electronic state approximations and calculations of observables. Calculations in \textsf{EXCEED-DM} are functions of the transition matrix elements, and the transition matrix element is only a function of the electronic configuration. In Sec.~\ref{subsec:calculations} we overview the calculations that can be performed in \textsf{v1.0.0} of \textsf{EXCEED-DM}. In Sec.~\ref{sec:inputs} we discuss the input files needed to perform a calculation with \textsf{EXCEED-DM}. In Sec.~\ref{sec:apps} we perform four new calculations with \textsf{EXCEED-DM} to illustrate what it can do, and how it can be used for publication quality results. Lastly, in Sec.~\ref{sec:summary} we discuss future extensions to \textsf{EXCEED-DM}.

\section{Formalism}
\label{sec:formalism}

\textsf{EXCEED-DM} is built to compute DM-electron interaction rates, e.g., DM scattering and absorption rates. While the formalism of DM-electron interactions has been previously discussed many times~\cite{Trickle:2019nya,Chen2022,Derevianko:2010kz,Mitridate:2021ctr,Gelmini:2020xir,Knapen:2021run,Hochberg:2021pkt,Hochberg:2017wce,Essig:2015cda,Graham:2012su,Geilhufe:2019ndy,Hochberg:2016sqx,Catena:2021qsr,Catena2022}, rarely is there a unifying theme to these derivations; each is usually performed for a specific type of interaction, or approximation concerning the electronic states. However, since the purpose of \textsf{EXCEED-DM} is to compute any of these interaction rates, the building blocks must be abstracted away from a specific calculation. In this section we will introduce these building blocks abstractly, and then discuss the explicit realizations which are currently implemented in \textsf{EXCEED-DM}. This should provide a solid foundation for future extensions relying on the same abstract building blocks, but with different realizations. 

\subsection{Electronic State Configuration}
\label{subsec:elec_state_config}

The principal building block is the electronic state, represented as $| I \rangle$. These are unit normalized, $\langle I | I \rangle  = 1$, eigenstates of the target Hamiltonian, $\hat{H}_0$, 
\begin{align}
    \hat{H}_0 | I \rangle = E_I | I \rangle \, .
\end{align}
A target is then nothing more than a collection of (orthonormal) electronic states $\{ | I \rangle \}$. Within a target we can define two types of these electronic states: initial, those that are initially filled and final, those that are empty and may be transitioned to. An \textit{electronic state configuration}, or \textit{electronic configuration}, is simply a collection of initial and final electronic states, $\{ \{ | I \rangle \}, \{ | F \rangle \} \}$. 

Clearly, an electronic state configuration depends on the target material, e.g., a Si target will contain a different set of initial and final electronic states than a Ge target. An electronic configuration file is the main target dependent input to \textsf{EXCEED-DM}.\footnote{See the documentation~\DocLink for more detailed information on explicitly constructing the electronic configuration file.} Moreover, this is where approximations about the target can enter; a different combination of electronic states constitutes a different approximation about the target. Both initial and final electronic states in the electronic configuration file may be chosen from the available electronic state approximations.

Currently, the available electronic state approximations are specific to crystal targets, i.e., those with a periodic potential. By Bloch's theorem these states can be indexed by a band number, $i$, and a momentum vector, $\mathbf{k}$, which lies inside the first Brillouin zone (1BZ). We will refer to these states specifically as \textit{Bloch states} throughout. The position space representation of these states is
\begin{align}
    \langle \mathbf{x}, s | i, \mathbf{k} \rangle =  e^{i \mathbf{k} \cdot \mathbf{x}} u_{i, \mathbf{k}, s}(\mathbf{x})  \, 
\end{align}
where $s$ is the component of the spin in the $\hat{\mathbf{z}}$ direction, and $u_{i, \mathbf{k}, s}$ are the, periodic and dimensionless, \textit{Bloch wave functions}, 
\begin{align}
    u_{i, \mathbf{k}, s}(\mathbf{x} + \mathbf{r}) = u_{i, \mathbf{k}, s}(\mathbf{x}) \, ,
\end{align}
where $\mathbf{r}$ is a lattice vector. Therefore a Bloch state can be represented by a function which returns $u_{i, \mathbf{k}, s}$. Note that this allows for another layer of abstraction; to implement other approximations for Bloch states one only needs to extend the Bloch wave function implementation with methods for computing $u_{i, \mathbf{k}, s}$. There are three different types of Bloch states which can currently be used in an electronic configuration file, each corresponding to a different basis: \textsf{bloch\_PW\_basis}, \textsf{bloch\_STO\_basis}, and \textsf{bloch\_single\_PW}. Different coefficients must be supplied for each basis. 

The \textsf{bloch\_single\_PW} are simply plane waves, 
\begin{align}
    u_{\mathbf{p}, \mathbf{k}(\mathbf{p})}(\mathbf{x}) = e^{i \left( \mathbf{p} - \mathbf{k}(\mathbf{p}) \right) \cdot \mathbf{x} }
\end{align}
which only requires specifying a momentum vector, $\mathbf{p}$, for which the corresponding $\mathbf{k} \in $ 1BZ is determined.\footnote{When a \textsf{bloch\_single\_PW} state is included as a final state, a Fermi factor, $F$, \begin{align}F(\nu) = \frac{\nu}{1 - e^{-\nu}} \\ \nu(Z_\text{eff}, E) = 2 \pi Z_\text{eff} \frac{\alpha m_e}{\sqrt{2 m_e E}} \, , \end{align} is used ($| F \rangle \rightarrow \sqrt{F(\nu)} | F \rangle$), to account for the fact that the final state wave function is not completely free. See Refs.~\cite{Lee:2015qva, Essig:2015cda, Graham:2012su, DarkSide:2018bpj, Griffin:2021znd} for more details.}

In the \textsf{bloch\_PW\_basis} basis, the coefficients, $\widetilde{u}_{i, \mathbf{k}, s, \mathbf{G}}$ must be specified to define the Bloch wave function,
\begin{align}
    u_{i, \mathbf{k}, s}(\mathbf{x}) = \sum_\mathbf{G} e^{i \mathbf{G} \cdot \mathbf{x}} \widetilde{u}_{i, \mathbf{k}, s, \mathbf{G}} \, .
\end{align}

In the \textsf{bloch\_STO\_basis} the Bloch wave functions are a linear combination of Slater type orbitals (STOs)~\cite{Bunge:1993jsz,Belkic1989} of each atom in the unit cell. The bands are indexed by the principal quantum numbers, $n, \ell, m$, as well at the index of the atom in the unit cell, $\kappa$,
\begin{align}
    u_{\kappa, n, \ell, m, \mathbf{k}}(\mathbf{x}) & = \sqrt{\Omega} \sum_{\mathbf{r}} e^{- i \mathbf{k} \cdot \mathbf{y}_{\mathbf{r}, \kappa}} \sum_j C_{j, \ell, n, \kappa} R_\text{STO}(y_{\mathbf{r}; \kappa}, Z_{j, \ell, \kappa}, n_{j, \ell, \kappa}) Y_l^m(\hat{\mathbf{y}}_{\mathbf{r}, \kappa})\, , \\
    R_\text{STO}(r; Z, n) & = a_0^{-3/2} \frac{(2Z)^{n + \frac{1}{2}}}{\sqrt{(2n)!}} \left( \frac{r}{a_0} \right)^{n - 1} e^{-Zr / a_0}
\end{align}
where $\Omega$ is the size of the unit cell, $\mathbf{y}_{\mathbf{r}, \kappa} \equiv \mathbf{x} - \mathbf{r} - \mathbf{x}_\kappa$, $\mathbf{x}_\kappa$ is the position of the atom in the unit cell.\footnote{In theory the sum over lattice vectors $\mathbf{r}$ should extend to infinity, however since the atomic wave functions are rapidly decaying, it is usually a good approximation to keep only the nearest neighboring cells.} Therefore to specify a \textsf{bloch\_STO\_basis} state, the central ingredients are the STO coefficients $\{ C_{j, \ell, \kappa}, Z_{j, \ell, \kappa}, n_{j, \ell, \kappa} \}$. A standard resource for these coefficients are tabulated for $2 \leq Z \leq 54$ at Ref.~\cite{Bunge:1993jsz}. They are also included in the \textsf{utilities/RHF\_wf\_data/RHF\_wf\_data.hdf5} file in the \textsf{EXCEED-DM} repository~\GithubLink for convenience. 

\subsection{Transition Matrix Elements}
\label{subsec:transition_matrix_elements}

Generally, DM-electron interaction rates cause an electron from an initial state to transition to a, higher energy, final state. Fundamentally this means that the most important quantity to compute are matrix elements between initial and final states, or \textit{transition matrix elements},
\begin{align}
    \mathcal{T}_{IF}(\hat{\mathcal{O}}) = \langle F | \mathcal{O} | I \rangle \, ,
\end{align}
where $\mathcal{O}$ is some quantum mechanical operator generating the transition between initial state $| I \rangle$ and final state $| F \rangle$. These transition matrix elements are the basis of \textsf{EXCEED-DM}. Given an electronic configuration, \textsf{EXCEED-DM} computes the resulting transition matrix elements, in parallel, which are then used in calculations of a specific observable, e.g., DM-electron scattering. The connection between the transition matrix elements and available calculations is discussed in Sec.~\ref{subsec:calculations}. This places the transition matrix elements directly between the approximations concerning the electronic states in the target, and those which concern the observable. Moreover, any observable which can be written in terms of the transition matrix elements can, in theory, be computed with \textsf{EXCEED-DM}.   

Currently, there are six transition form factors which can be computed,
\begin{align}
    \mathcal{T}_1(\mathbf{q}) & \equiv \langle F | e^{i \mathbf{q} \cdot \mathbf{x}} | I \rangle \label{eq:T_1}\\
    \mathcal{T}_\mathbf{v}(\mathbf{q}) & \equiv \langle F | e^{i \mathbf{q} \cdot \mathbf{x}} \, \mathbf{v} | I \rangle \label{eq:T_v} \\
    \mathcal{T}_{v^2}(\mathbf{q}) & \equiv \langle F | e^{i \mathbf{q} \cdot \mathbf{x}} \, v^2 | I \rangle \label{eq:T_v2} \\
    \mathcal{T}_{\bm{\sigma}}(\mathbf{q}) & \equiv \langle F | e^{i \mathbf{q} \cdot \mathbf{x}} \, \bm{\sigma} | I \rangle \label{eq:T_s} \\
    \mathcal{T}_{\mathbf{v} \cdot \bm{\sigma}}(\mathbf{q}) & \equiv \langle F | e^{i \mathbf{q} \cdot \mathbf{x}} \, \left( \mathbf{v} \cdot \bm{\sigma} \right) | I \rangle \label{eq:T_vds} \\
    \mathcal{T}_{\mathbf{v} \times \bm{\sigma}}(\mathbf{q}) & \equiv \langle F | e^{i \mathbf{q} \cdot \mathbf{x}} \, \left( \mathbf{v} \times \bm{\sigma} \right) | I \rangle \label{eq:T_vxs} \, ,
\end{align}
where $\mathbf{v} \equiv \mathbf{p}_e/m_e$, $\mathbf{p}_e$ is the electron momentum operator, $\bm{\sigma}$ are the Pauli matrices, and we have left the $I, F$ implicit, for readability. Additionally, each of these operators can be computed in the $\mathbf{q} \rightarrow 0$ limit where the exponential factor, $e^{i \mathbf{q} \cdot \mathbf{x}}$, is dropped. Typically, the $\mathbf{q} \rightarrow 0$ limit operators are only used in absorption calculations.

For each electronic state approximation, new routines must be added to compute these $\mathcal{T}$. For example, for the Bloch states discussed in Sec.~\ref{subsec:elec_state_config}, $\mathcal{T}_1(\mathbf{q})$, between an initial state indexed by $I = \{ i, \mathbf{k} \}$, and a final state indexed by $F = \{ f, \mathbf{k}' \}$, is given by,
\begin{align}
    \mathcal{T}^1_{i,f,\mathbf{k},\mathbf{k}'}(\mathbf{q} = \mathbf{k}' - \mathbf{k} + \mathbf{G}) & = \frac{1}{\Omega} \sum_s \int_\text{UC} d^3\mathbf{x} \, e^{i \mathbf{G} \cdot \mathbf{x}}u^*_{f, \mathbf{k}', s}  u_{i, \mathbf{k}, s} \, ,
\end{align}
where $\Omega$ is the volume of the unit cell (UC). See App.~\ref{app:T_for_bloch} for the derivation and general form of $\mathcal{T}$ for the other operators in Eqs.~\eqref{eq:T_1} -~\eqref{eq:T_vxs}.

\subsection{Calculations}
\label{subsec:calculations}

With the electronic state configuration and transition matrix elements defined, any DM-electron interaction rate written in terms of these quantities can, in principle, be added to \textsf{EXCEED-DM} and computed. Currently there are three calculation types supported, \textsf{binned\_scatter\_rate}, \textsf{absorption\_rate}, and \textsf{dielectric} which are discussed in detail below. While each of these calculations computes a specific quantity, other variables in the input file, see Sec.~\ref{sec:inputs}, can be set to specify different parameters to be used in the calculation. For example, the DM-electron scattering rate computed inside \textsf{binned\_scatter\_rate} can account for a variety of scattering form factors, which is useful when studying a variety of DM models. The purpose of this section is to give the explicit, general formulas that each calculation is computing, to understand the scope of each calculation. We will not give derivations for each of these formulae, and encourage the reader to follow the references for more details about individual calculations.

\subsubsection{Binned Scatter Rate}

\renewcommand{\arraystretch}{1.2}
\setlength{\tabcolsep}{12pt}
\begin{table}
    \begin{center}
        \begin{tabular}{@{}cc@{}} \toprule
            \textsf{FIF\_id} & $\mathscr{F}_{IF}$ \\ \midrule
            \textsf{'SI'} & $|\mathcal{T}_{1}|^2$ \\
            \textsf{'SD'} & $\displaystyle \frac{1}{3} |\mathcal{T}_{\bm{\sigma}}|^2$ \\
            \textsf{'VA1'} & $\displaystyle \frac{1}{\alpha^2 m_e^2} \left( 4 m_e^2 |\mathcal{T}_\mathbf{v}|^2 + 2 m_e \mathcal{T}_1 (\mathbf{q} \cdot {\mathcal{T}_\mathbf{v}}^*) + 2 m_e {\mathcal{T}_1}^* (\mathbf{q} \cdot {\mathcal{T}_\mathbf{v}}) + q^2 |\mathcal{T}_1|^2  \right)$ \\ \bottomrule
        \end{tabular}
    \end{center}
    \caption{Scattering form factors, $\mathscr{F}_{IF}$, currently provided with \textsf{EXCEED-DM}, written in terms of the transition matrix elements, $\mathcal{T}$, in Eqs.~\eqref{eq:T_1} -~\eqref{eq:T_vxs}. Different scattering form factors can be specified by changing the \textsf{FIF\_id} flag in the \textsf{dm\_model} input group discussed in Sec.~\ref{sec:inputs}. Note the `SD' scattering form factor is only available for two component wave functions; for spin degenerate electronic states it becomes $\propto |\mathcal{T}_1|^2$.} 
    \label{tab:binned_scatter_rate_table}
\end{table}
\renewcommand{\arraystretch}{1}

We begin by discussing the DM-electron scattering rate calculation. A single formula for DM-electron scattering, per kg-year exposure, $R$, derived in a variety of other papers~\cite{Trickle:2019nya,Chen2022,Knapen:2021run,Hochberg:2021pkt,Essig:2015cda,Graham:2012su,Catena:2021qsr,Catena2022}, encompasses a wide number of physical processes. Assuming the scattering rate does not depend on the DM velocity, $R$ is given by, 
\begin{align}
    R & = \frac{\pi \bar{\sigma}_e}{V \mu_{\chi e}^2 m_\chi} \frac{\rho_\chi}{\rho_T}  \sum_{I F} \int \frac{d^3 \mathbf{q}}{(2\pi)^3} f_\text{scr}^2 \mathcal{F}_\text{med}^2 g(\mathbf{q}, E_F - E_I) \mathscr{F}_{IF}(\mathcal{T}) \, ,
    \label{eq:scatter_rate}
\end{align}
where $\bar{\sigma}_e$ is a reference cross section, dependent on the couplings in the UV Lagrangian, $m_\chi$ is the DM mass, $\mu_{\chi e}$ is the DM-electron reduced mass, $\rho_\chi$ is the DM density, $\rho_T$ is the target density, $V$ is the target volume, $I, F$ index the initial and final electronic states in the target and $\mathbf{q}$ is the momentum transferred to the target. We will now go through each of the terms in the integrand in detail. $g(\mathbf{q}, \omega)$ is the \textit{kinematic function}, encoding the DM velocity distribution dependence,
\begin{align}
    g(\mathbf{q}, \omega) = 2 \pi \int d^3\mathbf{v} f_\chi(\mathbf{v}) \, \delta\left( \omega - \mathbf{q} \cdot \mathbf{v} + \frac{q^2}{2 m_\chi}\right) \, ,
\end{align}
where $f_\chi(\mathbf{v})$ is the DM velocity distribution, in the lab frame. A standard choice for velocity distribution is the Standard Halo Model (SHM)~\cite{Evans2018,Baxter:2021pqo} which has three parameters, the velocity spread, $v_0$, the galactic escape velocity, $v_\text{esc}$, and the Earth velocity relative to the galactic center, $\mathbf{v}_e$. The kinematic function for the SHM model is known analytically,
\begin{align}
    g(\mathbf{q}, \omega) & = \frac{2 \pi^2 v_0^2}{q N_0} \left(e^{-v_-^2/v_0^2} - e^{-v_\text{esc}^2/v_0^2} \right)   \label{eq:kin_func} \\
    v_- & = \text{min} \left\{ \frac{1}{q} \left| \, \omega + \frac{q^2}{2m_\chi} + \mathbf{q} \cdot \mathbf{v}_e \right|, v_\text{esc} \right\} \\
    N_0 & = \pi^{3/2} v_0^2 \left[ v_0 \text{erf} \left( v_\text{esc}/v_0 \right) - \frac{2 v_\text{esc}}{\sqrt{\pi}} \text{exp}\left( - v_\text{esc}^2/v_0^2 \right) \right] \, ,
\end{align}
where $\text{erf}$ is the error function.\footnote{While \textsf{EXCEED-DM} defaults to using the SHM (with user specified values of the velocity parameters), other velocity distributions may be readily added in the future, given a new kinematic function, $g$, Eq.~\eqref{eq:kin_func}, is provided.} Annual and daily modulation effects, e.g., the daily modulation rate in an anisotropic target, considered in Ref.~\cite{Trickle:2019nya}, enter the calculation through specific choices of $\mathbf{v}_e(t)$.

$f_\text{scr}$ and $\mathcal{F}_\text{med}$ in Eq.~\eqref{eq:scatter_rate} parameterize the propagator dependence. $f_\text{scr}$ incorporates screening effects due to mixing between the propagating dark particle, e.g., a dark photon, and the photon. In the absence of screening effects, $f_\text{scr} = 1$, but different DM models can introduce different screening factors. The primary example where screening is important is in a model with a dark photon coupling to the electron vector current, analogous to the Standard Model photon. In this case, the screening factor is dependent on the dielectric function,
\begin{align}
    f_\text{scr} = \frac{1}{\hat{\mathbf{q}} \cdot \bm{\varepsilon}(\mathbf{q}, \omega) \cdot \hat{\mathbf{q}}} \, . \label{eq:f_scr}
\end{align}
\textsf{EXCEED-DM} allows the user to set $f_\text{scr}$ to $1$, use an analytic model for $\varepsilon(q, \omega)$~\cite{Cappellini1993}, or use a precomputed dielectric function in an external file. $\mathcal{F}_\text{med}$ is the mediator propagator, not including screening effects, parameterized by $\beta$,
\begin{align}
    \mathcal{F}_\text{med} = \left( \frac{\alpha m_e}{q} \right)^\beta \, ,
    \label{eq:med_ff}
\end{align}
where $\beta = 0$ for a heavy mediator, and $\beta = 2$ for a light mediator.

The last term to discuss in Eq.~\eqref{eq:scatter_rate} is the scattering form factor, $\mathscr{F}_{IF}$. This is intrinsically DM model dependent, and will take different forms depending on the DM model. Table~\ref{tab:abs_rate_table} contains the currently supported options in \textsf{EXCEED-DM}, useful for a range of DM models. See Appendix~\ref{app:eff_L_to_F} for a (partial) derivation from a general UV Lagrangian to $\mathscr{F}_{IF}$ in terms of the transition matrix elements. Adding new $\mathscr{F}_{IF}$ can be done with relative ease within the \textsf{EXCEED-DM} codebase since $\mathscr{F}_{IF}$ need only be a function of the transition matrix elements; one does not have to worry about deriving the relevant formulas in terms of electronic wave functions, which may be very different depending on the initial and final state approximations. 

In addition to the total scattering rate given in Eq.~\eqref{eq:scatter_rate}, \textsf{EXCEED-DM} computes the scattering rate binned in energy and momentum deposition, i.e., the \textit{binned scatter rate}. This is a matrix, $R_{ij}$, whose $(i, j)$ component contains the total event rate for events with $\omega, q$ in 
\begin{align}
    & (i - 1) \times \Delta \omega < \omega - E_g < i \times \Delta \omega \\
    & (j - 1) \times \Delta q < q < j \times \Delta q
\end{align}
where $\Delta \omega, \Delta q$ are the bin widths, $E_g$ is the band gap, and we assume the first index is one. $R_{ij}$ is defined such that $R = \sum_{ij} R_{ij}$, for any choice of $\Delta \omega, \Delta q$.\footnote{The last bin, i.e., $i = N_\omega$ or $j = N_q$, is special. It contains the event rate for events with $\omega > (N_\omega - 1) \times \Delta \omega$, and $q > (N_q - 1) \times \Delta q$. This is to ensure $\sum_{ij} R_{ij} = R$ is satisfied for any choice of $N_\omega, N_q, \Delta \omega, \Delta q$.}

\subsubsection{Absorption Rate}

\renewcommand{\arraystretch}{1.2}
\setlength{\tabcolsep}{12pt}
\begin{table}
    \begin{center}
        \begin{tabular}{@{}cccc@{}} \toprule
            \textsf{particle\_id} & $\delta \mathcal{L}$ & Wave Function Type & $\Pi_{\hat{\phi} \hat{\phi}}^\eta$ \\ \midrule
            \multirow{2}{*}{ \textsf{'scalar'} } & \multirow{2}{*}{ $g_e \phi \bar{e} e$ } & \multirow{2}{*}{SI/SD} & \multirow{2}{*}{$\displaystyle \frac{g_e^2}{4} \bar{\Pi}_{v^2, v^2}$} \\ & & &  \\ 
            \multirow{2}{*}{ \textsf{'ps'} } & \multirow{2}{*}{ $i g_e \phi \bar{e} \gamma^5 e$ } & SI & $\displaystyle g_e^2 \frac{m_\phi^2}{4 m_e^2} \sum_i \bar{\Pi}_{v^i, v^i}$ \\ & & SD & $\displaystyle g_e^2 \frac{m_\phi^2}{4 m_e^2} \bar{\Pi}_{\mathbf{v} \cdot \bm{\sigma}, \mathbf{v} \cdot \bm{\sigma}}$ \\ 
            \multirow{2}{*}{ \textsf{'vector'} } & \multirow{2}{*}{ $g_e \phi_\mu \bar{e} \gamma^\mu e$ } & \multirow{2}{*}{SI/SD} & \multirow{2}{*}{$\displaystyle g_e^2 \frac{ m_\phi^2 \left[ \bar{\Pi}_{\mathbf{v}, \mathbf{v}}' \right]^\eta}{ m_\phi^2 - e^2 \left[ \bar{\Pi}_{\mathbf{v}, \mathbf{v}}' \right]^\eta}$} \\ & & & \\ 
        \bottomrule
        \end{tabular}
    \end{center}
    \caption{Absorption rate calculations available in \textsf{EXCEED-DM}. Specify a model with the \textsf{particle\_id} variable in the \textsf{dm\_model} input group. Each model corresponds to an interaction Lagrangian, $\delta \mathcal{L}$. The pseudoscalar DM calculation is different depending on whether the input wave functions are one (``SI") or two (``SD") dimensional. $\bar{\Pi}_{\mathcal{O}_1, \mathcal{O}_2}$ and $\bar{\Pi}'_{\mathcal{O}_1, \mathcal{O}_2}$ are defined in Eq.~\eqref{eq:bar_Pi} and Eq.~\eqref{eq:bar_Pi_prime}, respectively. $\left[ \bar{\Pi}'_{\mathbf{v}, \mathbf{v}}\right]^\eta$ is shorthand for an eigenvalue of the $3\times 3$ matrix, $\bar{\Pi}_{\mathbf{v}, \mathbf{v}}'$.} 
    \label{tab:abs_rate_table}
\end{table}
\renewcommand{\arraystretch}{1}

We now turn to the absorption rate calculation. Similar to the DM-electron scattering rate calculation, the formalism has been studied in detail previously~\cite{Chen2022,Derevianko:2010kz,Mitridate:2021ctr,Gelmini:2020xir,Hochberg:2016sqx}. The general absorption rate, $R$, per kg-year, of DM particle $\phi$ is given by,
\begin{align}
    R & = -\frac{\rho_\phi}{\rho_T m_\phi^2} \frac{1}{n} \sum_{\lambda = 1}^{n} \text{Im} \left[ \Pi_{\hat{\phi} \hat{\phi}}^\lambda(\bar{\Pi}_{\mathcal{O}_1, \mathcal{O}_2}) \right] \nonumber \\
      & =-\frac{\rho_\phi}{\rho_T m_\phi^2} \frac{1}{n} \sum_{\lambda = 1}^{n} \text{Im} \left[ \Pi_{\phi\phi}^\lambda + \sum_\eta \frac{\Pi_{\phi A}^{\lambda \eta} \Pi_{A\phi}^{\eta \lambda}}{m_\phi^2 - \Pi_{AA}^\eta} \right] \label{eq:abs_rate}
\end{align}
where $\rho_\phi$ is the DM density, $m_\phi$ is the DM mass, $n$ is the number of degrees of freedom of $\phi$, $\Pi_{\hat{\phi} \hat{\phi}}^\lambda$ is the self-energy of the $\phi$-like state in medium, projected on to the $\lambda$th polarization. The difference between $\phi$ and the $\phi$-like state is due to mixing effects between $\phi$ and $A$. $\Pi_{\phi_1 \phi_2}^{\lambda_1 \lambda_2}$ is the self-energy between the $\lambda_1$ polarization of $\phi_1$, and the $\lambda_2$ polarization of $\phi_2$ (when $\phi_1 = \phi_2$, we simplify the notation by only having one polarization index, $\lambda_1 = \lambda_2 = \lambda$). 

The self-energies in Eq.~\eqref{eq:abs_rate} can be written in terms of the transition form factors in the $\mathbf{q} \rightarrow 0$ limit. To simplify this we introduce $\bar{\Pi}_{\mathcal{O}_1, \mathcal{O}_2}$ which is a (polarization independent) self-energy computed from a one loop diagram~\cite{Chen2022,Mitridate:2021ctr},
\begin{align}
\bar{\Pi}_{\mathcal{O}_1, \mathcal{O}_2} & \equiv \frac{1}{V} \sum_{IF} G(\omega, E_F - E_I, \delta) \mathcal{T}_{\mathcal{O}_1}(\mathbf{q} \rightarrow 0) \mathcal{T}^*_{\mathcal{O}_2}(\mathbf{q} \rightarrow 0) \label{eq:bar_Pi} \\
\bar{\Pi}'_{\mathcal{O}_1, \mathcal{O}_2} & \equiv \frac{1}{V} \sum_{IF} G(\omega, E_F - E_I, \delta) \left( \frac{\omega}{E_F - E_I} \right)^2 \mathcal{T}_{\mathcal{O}_1}(\mathbf{q} \rightarrow 0) \mathcal{T}^*_{\mathcal{O}_2}(\mathbf{q} \rightarrow 0) \label{eq:bar_Pi_prime}\\
G(\omega, \Delta \omega, \delta) & = \frac{1}{\omega - \Delta \omega + i \delta} - \frac{1}{\omega + \Delta \omega - i \delta} \label{eq:G_eq}\, ,
\end{align}
where $\delta$ is the electron lifetime (width).\footnote{We also introduce $\bar{\Pi}'$, which alters the $\omega$ scaling relative to $\bar{\Pi}$, to avoid computing terms which can analytically be shown to be zero. See Ref.~\cite{Sangalli2017} for a more detailed discussion. Note that in the $\delta \rightarrow 0$ limit, $\text{Im} \left[ \bar{\Pi} \right] = \text{Im} \left[ \bar{\Pi}' \right]$, and therefore a converged calculation should be identical when computed with $\bar{\Pi}'$ or $\bar{\Pi}$.} Table~\ref{tab:abs_rate_table} lists the, currently supported, DM model dependent expressions for $\Pi_{\hat{\phi}\hat{\phi}}$, in Eq.~\eqref{eq:abs_rate}, in terms of the $\bar{\Pi}$'s, as derived in Refs.~\cite{Chen2022, Mitridate:2021ctr}.

\subsubsection{Dielectric Function}
\label{subsubsec:dielectric}

The last calculation included in \textsf{EXCEED-DMv1.0.0} is of the dielectric function, $\varepsilon(\mathbf{q}, \omega)$, which can be computed from the Lindhard formula~\cite{Ashcroft1976}, and written in terms of $\mathcal{T}_1$ as, 
\begin{align}
    \varepsilon(\mathbf{q}, \omega) & = 1 - \frac{e^2}{q^2 V} \sum_{IF} G(\omega, E_F - E_I, \delta) |\mathcal{T}_1|^2 \, , \label{eq:dielectric}
\end{align}
where $G$ is defined in Eq.~\eqref{eq:G_eq}, $e$ is the electromagnetic charge, and $V$ is the target volume. We include this calculation in \textsf{EXCEED-DM} for two reasons. First, to provide a consistent calculation of DM-electron scattering rates, the screening factor in Eq.~\eqref{eq:f_scr} should be computed under the same assumptions as the scattering form factor, $\mathscr{F}_{IF}$. This was not done in previous \textsf{EXCEED-DM} calculations for Si and Ge targets, e.g., in Ref.~\cite{Griffin:2021znd} which utilized an analytic dielectric function. Consistent calculations, performed with \textsf{v1.0.0} of \textsf{EXCEED-DM}, can be found in Sec.~\ref{subsec:dme_SI_numeric_dielectric}. Second, it has been shown that for a class of simple DM models, DM-electron interaction rates can be written solely in terms of the dielectric function~\cite{Knapen:2021run,Hochberg:2021pkt}. For these models, once a dielectric function has been computed, the package \textsf{DarkELF} can use these pre-computed dielectric functions and compute DM-electron interaction rates. Therefore adding the dielectric function as an output of \textsf{EXCEED-DM} is not only useful as a point of comparison to different calculations of the dielectric function, but also useful as an input to \textsf{DarkELF}.

Technically, the dielectric function in Eq.~\eqref{eq:dielectric} is not what is computed. While the dielectric function in Eq.~\eqref{eq:dielectric} is a function of $\mathbf{q}$, due to lattice momentum conservation it is only non-zero for $\mathbf{G} = \mathbf{q} - \mathbf{k}_F + \mathbf{k}_I$, where $\mathbf{k}_{I(F)}$ is the Bloch momentum of the initial (final) state, and $\mathbf{G}$ is a reciprocal lattice vector. Therefore, the requested $\mathbf{q}$ point may not be kinematically satisfied with the users choice of initial and final states. For example, if the datasets only contain $\mathbf{k}_I = \mathbf{k}_F = 0$, and $\mathbf{q}$ is not a reciprocal lattice vector, the result would be zero. We try to avoid constraining the user to pick an appropriate set of $\mathbf{q}$ points. The easiest way around this would be to keep track of all the $\mathbf{q}$ that \textit{are} included in the dataset, but an array of size $N_I \times N_F \times N_\mathbf{G}$ is generally too large to store.

To avoid these issues we compute an averaged dielectric function,
\begin{align}
\bar{\varepsilon}(\mathbf{q} \, \text{bin}, \omega) & \equiv \frac{1}{V_{\mathbf{q} \, \text{bin}}} \int_{\mathbf{q} \, \text{bin}}  \varepsilon(\mathbf{q}, \omega) \, d^3\mathbf{q} \\
                                                    & = 1 - \frac{e^2}{V V_{\mathbf{q} \, \text{bin}}} \int \frac{d^3 \mathbf{q}}{q^2} \sum_{IF} G(\omega, E_F - E_I, \delta) |\mathcal{T}_1|^2 \Theta( \mathbf{q} \in \mathbf{q} \, \text{bin} ) \, ,
\end{align}
where $V_{\mathbf{q} \, \text{bin}}$ is the volume of the bin around $\mathbf{q}$. This has the benefits of including all $\mathbf{q}$ points in the dataset, and the output is smaller in size since the sum is performed before saving the data. We compare this averaged dielectric function to the analytic model used previously~\cite{Griffin:2021znd} in Sec.~\ref{subsec:dme_SI_numeric_dielectric}. 

\section{Inputs to EXCEED-DM}
\label{sec:inputs}

Before discussing specific applications of \textsf{EXCEED-DM} in Sec.~\ref{sec:apps}, we will discuss the input files needed to run \textsf{EXCEED-DM}. For every calculation there are two necessary files: an \textsf{input} file containing a list of parameters to use for the calculation, and the \textsf{electronic configuration} file which contains all the information regarding the electronic configuration, discussed in more detail in Sec.~\ref{subsec:elec_state_config}. The discussion here regarding both of these input files is meant to serve as an introduction, so the setup of the calculations in Sec.~\ref{sec:apps} is clear. A complete list of input parameters, and their defaults, can be found in the documentation~\DocLink, as well as the full file specification for the \textsf{electronic configuration} file. A plethora of example \textsf{input} and \textsf{electronic configuration} files can be found in the \textsf{examples/} folder in the \textsf{EXCEED-DM} repository~\GithubLink.

\subsection{Input File}
\label{subsec:input_file}

The \textsf{input} file is just a text file where specific variables are specified in different \textit{groups}. Each group contains a collection of variables relevant to that group. The group is specified inside brackets, `[]', and variables are specified below it. For example, the \textsf{control} group, contains the variables, 
\begin{itemize}
    \item \textsf{calculation} - Specifying which calculation is performed.
    \item \textsf{out\_folder} - Specifying the folder where the output file should be saved.
    \item \textsf{run\_description} - A short description of the calculation, appended to the end of \textsf{'EXDM\_out\_'} when creating the output filename.
\end{itemize}
The corresponding \textsf{input} file would contain:
\begin{nice_box}{Input File}{BrickRed}
    [control]

        \begin{tabular}{ll}
            \inputtab calculation & = `binned\_scatter\_rate' \\
            \inputtab out\_folder & = `./examples/1/output/' \\
            \inputtab run\_description & = `example\_1'
        \end{tabular}
\end{nice_box}

While most of the input parameters take a specific value, some allow an array to be specified. For example, in the \textsf{dm\_model} group (containing information about the DM model to use in the calculation), any list of DM masses (in units of eV) can be specified in the \textsf{mX} variable,
\begin{nice_box}{Input File}{BrickRed}
    [dm\_model]

        \begin{tabular}{ll}
            \inputtab mX & = 1e6, 1e7, 1e8, 1e9, ...  \\
        \end{tabular}
\end{nice_box}
A list of variables can also be specified vertically with the \textsf{+=} operator, e.g.,
\begin{nice_box}{Input File}{BrickRed}
    [dm\_model]

        \begin{tabular}{ll}
            \inputtab mX & = 1e6 \\
            \inputtab mX & += 1e7 \\
            \inputtab mX & += 1e8 \\
            \inputtab mX & += 1e9 \\
                         &\vdots
        \end{tabular}
\end{nice_box}
Note that DM masses can also be specified with the \textsf{mX\_logspace}, \textsf{mX\_linspace} variables to add a logarithmically uniform, and uniform list of masses, respectively. See the documentation~\DocLink for more information about input file formatting specifics.

Currently, there are five main input groups which control a calculation: 
\begin{itemize}
    \item \textsf{control} - Controls what calculations are performed and where the output files are saved. 
    \item \textsf{elec\_config\_input} - Contains information about the \textsf{electronic configuration} input file. 
    \item \textsf{material} - Specifies some target material parameters.
    \item \textsf{dm\_model} - Sets the DM model parameters, e.g. DM masses, mediator type, etc.
    \item \textsf{astroph\_model} - Sets the astrophysical parameters of the DM velocity distribution.
\end{itemize}
For each calculation there is also a \textsf{numerics\_<calculation>} group which specifies some numerical parameters specific to that calculation, e.g., energy bin width in a scattering rate calculation. Having already discussed the \textsf{control} group above, we will now discuss, and give an example of each of these other groups.

The \textsf{elec\_config\_input} group is the simplest group, currently containing only a single variable, \textsf{filename}, which contains the path to the \textsf{electronic configuration} file,

\begin{nice_box}{Input File}{BrickRed}
    [elec\_config\_input]

        \begin{tabular}{ll}
            \inputtab filename = './examples/1/elec\_config.hdf5'
        \end{tabular}
\end{nice_box}

The \textsf{material} group, contains three main variables, \textsf{name} (the name of the material), \textsf{rho\_T\_g\_per\_cm3} (the target mass density in $\text{g}/\text{cm}^3$), and \textsf{a\_vecs\_Ang}, the three primitive lattice vectors, $\mathbf{a}_i$, in units of \AA,

\begin{nice_box}{Input File}{BrickRed}
    [material]

        \begin{tabular}{ll}
            \inputtab name &= `Si' \\
            \inputtab rho\_T\_g\_per\_cm3 &= 2.281 \\
            \inputtab \\ 
            \inputtab a\_vecs\_Ang &= 0.00000, 2.73437, 2.73437 \\
            \inputtab a\_vecs\_Ang &+= 2.73437, 0.00000, 2.73437 \\
            \inputtab a\_vecs\_Ang &+= 2.73437, 2.73437, 0.00000
        \end{tabular}
\end{nice_box}

An example \textsf{dm\_model} group is

\begin{nice_box}{Input File}{BrickRed}
    [dm\_model]

        \begin{tabular}{ll}
            \inputtab mX &= 1e6, 1e7, 1e8, 1e8, 1e10 \\
            \inputtab med\_FF &= 0, 2 \\
            \inputtab particle\_type & = `fermion' \\
            \inputtab FIF\_id &= `SI' \\
            \inputtab rho\_X\_GeV\_per\_cm3 &= 0.4
        \end{tabular}
\end{nice_box}
where \textsf{mX} are, again, the DM masses in eV, \textsf{med\_FF} is the power of the mediator form factor, $\beta$ in Eq.~\eqref{eq:med_ff}, (note that more than one may be specified at once), \textsf{particle\_type} is the type of DM particle, \textsf{FIF\_id} specifies the scattering form factor, and \textsf{rho\_X\_GeV\_per\_cm3} is the DM density in $\text{GeV}/\text{cm}^3$. Not all of these variables are used in every calculation. For example, \textsf{med\_FF} and \textsf{FIF\_id} are specific to the \textsf{binned\_scatter\_rate} calculation, and \textsf{particle\_type} is specific to the \textsf{absorption\_rate} calculation. 

Lastly, an example \textsf{astroph\_model} group is given by,
\begin{nice_box}{Input File}{BrickRed}
    [astroph\_model]

        \begin{tabular}{ll}
            \inputtab vel\_distribution\_name & = `SHM' \\
            \inputtab v\_0\_km\_per\_sec &= 230 \\
            \inputtab v\_esc\_km\_per\_sec &= 600 \\
            \\
            \inputtab v\_e\_km\_per\_sec &= 0, 0, 230 \\
            \inputtab v\_e\_km\_per\_sec &+= 0, 0, 240 \\
            \inputtab v\_e\_km\_per\_sec &+= 0, 0, 220 \\
        \end{tabular}
\end{nice_box}
where \textsf{vel\_distribution\_name} is the name of the velocity distribution used,\footnote{Currently, only `SHM' is available.} \textsf{v\_0\_km\_per\_sec} is the typical velocity parameter of the Standard Halo Model (SHM) in $\text{km}/\text{s}$, \textsf{v\_esc\_km\_per\_sec} is the galactic escape velocity at the location of the Earth in $\text{km}/\text{s}$, and \textsf{v\_e\_km\_per\_sec} is a list of Earth velocity vectors in the target crystal frame, i.e., the $\hat{\mathbf{x}}, \hat{\mathbf{y}}, \hat{\mathbf{z}}$ axes are taken to be the basis of the crystal lattice vectors, $\mathbf{a}_i$. Specifying a list of Earth velocity vectors allows daily and annual modulation signals to be calculated, see Sec.~\ref{subsec:annual_modulation} for a more detailed discussion of annual modulation, and Refs.~\cite{Trickle:2019nya,Blanco:2021hlm,Hochberg:2016ntt,Coskuner:2019odd,Coskuner:2019odd} for a more detailed discussion of daily modulation signals.  

\subsection{Electronic Configuration File}
\label{subsec:elec_config_file}

The electronic configuration file is stored in a hierarchical data format (\href{https://www.hdfgroup.org/solutions/hdf5/}{HDF5}). The pros of this approach are that the data is compressed, and easily (quickly) read in to \textsf{EXCEED-DM}. The cons are that the data is not easy to edit or view without external programs. We recommend the use of \href{https://www.hdfgroup.org/downloads/hdfview/}{\textsf{HDFView}} and the \textsf{python} program \href{https://www.h5py.org/}{\textsf{h5py}} for viewing and manipulating these files. In this section we will discuss the structure of these files, and leave specific details to the documentation~\DocLink. This should make clear where future extensions may add their components, and aid in understanding the example \textsf{electronic configuration} files, \textsf{elec\_config.hdf5}, in each example in the \textsf{examples/} folder on the Github page~\GithubLink. In the directory trees below, folders end with a backslash, and datasets do not. Extra folders/datasets may be included, but they will not be used in the calculations (without modification to \textsf{EXCEED-DM}). This is useful mainly to store explanatory metadata with the file. 

At the highest level there is simply an \textsf{elec\_states} folder with two similar subfolders, \textsf{init}, and \textsf{fin},

\begin{nice_box}{Electronic Configuration File}{NavyBlue}
    \renewcommand*\DTstyle{\sffamily}
    \dirtree{%
        .1 elec\_states/.
        .2 init/.
        .2 fin/.
    }
\end{nice_box}
\textsf{elec\_states} contains the data about the electronic states, \textsf{init} holds the data about the initial (filled) electronic states, and \textsf{fin} holds the data regarding the final (unfilled) electronic states. The electronic state information inside the \textsf{init}, \textsf{fin} folders should be grouped by level of approximations, i.e., all the data in a subfolder should correspond to electronic states defined in that approximation. For example, all electronic states under the ``Bloch" approximation, discussed in Sec.~\ref{subsec:elec_state_config}, should be inside a \textsf{bloch} folder:

\begin{nice_box}{Electronic Configuration File}{NavyBlue}
    \renewcommand*\DTstyle{\sffamily}
    \dirtree{%
        .1 elec\_states/.
        .2 init/.
        .3 bloch/.
        .2 fin/.
        .3 bloch/.
    }
\end{nice_box}
and different approximations should go in their own folder. For the Bloch states there are further approximations, specifically what basis the states are in, e.g.,

\begin{nice_box}{Electronic Configuration File}{NavyBlue}
    \renewcommand*\DTstyle{\sffamily}
    \dirtree{%
        .1 init/.
        .2 bloch/.
        .3 PW\_basis/.
        .3 STO\_basis/.
        .3 single\_PW/.
    }
\end{nice_box}
and similarly for the \textsf{fin} states. Within each of these approximation subfolders there should be two folders containing the actual datasets, \textsf{config}, and \textsf{state\_info}, e.g.,

\begin{nice_box}{Electronic Configuration File}{NavyBlue}
    \renewcommand*\DTstyle{\sffamily}
    \dirtree{%
        .1 bloch/.
        .2 PW\_basis/.
        .3 config/.
        .3 state\_info/.
        .2 STO\_basis/.
        .3 config.
        .3 state\_info/.
        .2 single\_PW/.
        .3 config/.
        .3 state\_info/.
    }
\end{nice_box}

The \textsf{config} folder is meant to store data common among all of the states of the same approximation, whereas the \textsf{state\_info} folder is meant to store arrays of data specific to each individual electronic state. For example, for the \textsf{bloch/PW\_basis} states,

\begin{nice_box}{Electronic Configuration File}{NavyBlue}
    \renewcommand*\DTstyle{\sffamily}
    \dirtree{%
        .1 PW\_basis/.
        .2 config/.
        .3 G\_red\_list\DTcomment{List of (reduced) $\mathbf{G}$ vectors each state is expanded for.}.
        .2 state\_info/.
        .3 Zeff\_list\DTcomment{$Z_\text{eff}$ for each state}.
        .3 energy\_list\DTcomment{$\omega$ for each state}. 
        .3 i\_list\DTcomment{Band index, $i$, for each state}.
        .3 jac\_list\DTcomment{Jacobian in $(1/N)\sum_I$ for each state (see below)}.
        .3 k\_id\_list\DTcomment{$\mathbf{k}$ id for each state (see below)}.
        .3 k\_vec\_red\_list\DTcomment{(reduced) $\mathbf{k}$ for each state}.
        .3 u\_FT\_c/\DTcomment{Complex component of the Bloch coefficients, $\widetilde{u}_\mathbf{G}$}.
        .4 n\_1\DTcomment{Complex component of the Bloch coefficients, $\widetilde{u}_\mathbf{G}$, for state $n$}.
        .4 \vdots.
        .3 u\_FT\_r/\DTcomment{Real component of the Bloch coefficients, $\widetilde{u}_\mathbf{G}$}.
        .4 n\_1\DTcomment{Real component of the Bloch coefficients, $\widetilde{u}_\mathbf{G}$, for state $n$}.
        .4 \vdots.
    }
\end{nice_box}

Other bases/approximations will have a different set of variables, see the documentation~\DocLink for the currently supported options. While most of the above entries should be clear, there are a few that deserve more clarification. The \textsf{k\_id\_list} array contains the index of that state's $\mathbf{k}$ vector, i.e., $i$ if the 1BZ vectors are indexed with $i$: $\mathbf{k}_i$. \textsf{jac\_list} represents the Jacobian of a given state when numerically computing the sum over the states, i.e., 
\begin{align}
    \frac{1}{N} \sum_I \rightarrow \sum_{i} j_i \, .
\end{align}
For example, consider of set of Bloch states, $I \in \{ i, \mathbf{k} \}$, sampled uniformly in the 1BZ with $N_\mathbf{k}$ $\mathbf{k}$ points, 
\begin{align}
    \frac{1}{N} \sum_I & = \frac{1}{N} \sum_i \sum_\mathbf{k} = \Omega \sum_i \int_\text{1BZ} \frac{d^3\mathbf{k}}{(2\pi)^3} \rightarrow \sum_i \sum_\mathbf{k} \frac{1}{N_\mathbf{k}} \\
    j_I = \frac{1}{N_\mathbf{k}} \, .
\end{align}
In this case the Jacobian was simple, however it need not be and will depend on the electronic states specified. Consider a set of \textsf{bloch/single\_PW}, $\{ \mathbf{p} \}$ states, sampled logarithmically in $E = p^2/2m_e$ between $E_\text{min}$ and $E_\text{max}$, and uniformly on the sphere in $(\theta_\mathbf{p}, \phi_\mathbf{p})$ space,
\begin{align}
    \frac{1}{N} \sum_I & = \Omega \int \frac{d^3 \mathbf{p}}{(2\pi)^3} = \frac{\Omega}{(2\pi)^3} \int dp \, p^2 \int d\theta \sin{\theta} \int d \phi \, .
    \label{eq:free_jac}
\end{align}
Uniformly sampling on the sphere involves changing variables to $\alpha = \phi/2\pi, \beta = (1/2) (1 + \cos{\theta})$, and sampling uniformly between $[ 0, 1 ]$ in $\alpha, \beta$. Sampling logarithmically in $E$ involves changing variables to $x = \ln{E}$, and sampling uniformly in $x$ between $[ \ln{E_\text{min}}, \ln{E_\text{max}} ]$. Therefore Eq.~\eqref{eq:free_jac} becomes
\begin{align}
    \frac{\Omega}{(2\pi)^3} \int dp \, p^2 \int d\theta \sin{\theta} \int d \phi & \rightarrow \frac{\Omega}{(2\pi)^3} \frac{2 \pi}{N_\alpha N_\beta N_x} \ln{\left( \frac{E_\text{max}}{E_\text{min}} \right)} \sum_{\alpha, \beta, x} p(x)^3 \\
    j_I & = \frac{\Omega}{(2\pi)^3} \frac{2 \pi}{N_\alpha N_\beta N_x}\ln{\left( \frac{E_\text{max}}{E_\text{min}} \right)} p(x)^3
\end{align}
where $I \in \{ \alpha, \beta, x \}$. We see that the Jacobian is no longer as simple. Since the Jacobian is inherently tied to the combination of states in the \textsf{electronic configuration}, we leave it to the user to take care of defining these properly.

\section{Applications}
\label{sec:apps}

Practically, the best way to understand what \textsf{EXCEED-DM} can do is to see its applications. The example calculations in the \textsf{examples} folder offer an introduction to these calculations with small datasets. Additionally, \textsf{EXCEED-DM} has been used previously in a number of papers, see Refs.~\cite{Trickle:2019nya,Zhang2022,Chen2022,Mitridate:2021ctr,Griffin:2021znd,Griffin:2019mvc,Dutta2022}. However, these papers have not gone in to detail on constructing the inputs for \textsf{EXCEED-DM}. Therefore the purpose of this section is two-fold: showcase the new additions to \textsf{v1.0.0} of \textsf{EXCEED-DM} by performing new calculations, and explicitly detailing the input files so it is clear how to perform these calculations for other target materials, DM models, etc.

Our focus will be on Si and Ge targets, used in SuperCDMS~\cite{SuperCDMS:2020ymb,SuperCDMS:2018mne,CDMS:2009fba}, DAMIC~\cite{DAMIC:2020cut,DAMIC:2016qck,Settimo:2020cbq,DAMIC:2015znm,DAMIC:2019dcn}, EDELWEISS~\cite{EDELWEISS:2018tde,EDELWEISS:2019vjv,EDELWEISS:2020fxc}, SENSEI~\cite{Crisler:2018gci,SENSEI:2020dpa,SENSEI:2019ibb}, and CDEX~\cite{Zhang2022}. The \textsf{electronic configuration} files are publicly available here~\ZenodoElecConfigLink. The initial states contain a combination of states in the \textsf{bloch\_STO\_basis} and \textsf{bloch\_PW\_basis}, and the final states contain a combination of states in the \textsf{bloch\_PW\_basis} and \textsf{bloch\_single\_PW} basis. In the language of Ref.~\cite{Griffin:2021znd}, these correspond to a set of ``core" and ``valence" initial states, and ``conduction" and ``free" final states. More specific details about the electronic states included in these files can be found in the documentation~\DocLink. All of the calculations shown below can be done with these \textsf{electronic configuration} files combined with the input groups shown in each section. The raw output files can be downloaded here~\ZenodoDataLink. To avoid repetition, the \textsf{material} group variables in the input file for Si and Ge are,
\begin{nice_box}{Input File}{BrickRed}
    [material]

        \begin{tabular}{ll}
            \inputtab name &= `Si' \\
            \inputtab rho\_T\_g\_per\_cm3 &= 2.281 \\
            \inputtab \\ 
            \inputtab a\_vecs\_Ang &= 0.00000, 2.73437, 2.73437 \\
            \inputtab a\_vecs\_Ang &+= 2.73437, 0.00000, 2.73437 \\
            \inputtab a\_vecs\_Ang &+= 2.73437, 2.73437, 0.00000
        \end{tabular}
\end{nice_box}

\begin{nice_box}{Input File}{BrickRed}
    [material]

        \begin{tabular}{ll}
            \inputtab name &= `Ge' \\
            \inputtab rho\_T\_g\_per\_cm3 &= 5.042 \\
            \inputtab \\ 
            \inputtab a\_vecs\_Ang &= 0.00000, 2.87981, 2.87981 \\
            \inputtab a\_vecs\_Ang &+= 2.87981, 0.00000, 2.87981 \\
            \inputtab a\_vecs\_Ang &+= 2.87981, 2.87981, 0.00000
        \end{tabular}
\end{nice_box}
respectively. These should be added to the calculation specific input groups discussed in each section below. We will also omit filename input variables which will be computer specific. Specifically, the \textsf{out\_folder}, and \textsf{run\_description} variables in the \textsf{control} group, and \textsf{filename} in the \textsf{elec\_config\_input} group. 

This section is organized as follows, in Sec.~\ref{subsec:dme_SI_numeric_dielectric} we will perform two new calculations. First, we will compute the average dielectric function, $\bar{\varepsilon}(q, \omega)$, with \textsf{EXCEED-DM} and compare it with the analytic dielectric function model used in Ref.~\cite{Sangalli2017} to screen the DM-electron interaction. Second, we will use this numerically computed dielectric function to screen the DM-electron scattering rate in the kinetically mixed dark photon model, and compare to the scattering rate with no screening, as well as the scattering rate from Ref.~\cite{Griffin:2021znd} which used the analytic dielectric function to screen the interactions. In Sec.~\ref{subsec:dme_VA} we study DM models beyond the simple dark photon model, and compute the DM-electron scattering rate in an example model which has a scattering potential dependent on the electron velocity. In Sec.~\ref{subsec:extended_absorption} we compute scalar, pseudoscalar, and vector DM absorption rates up to DM masses of a keV. This extends the calculation done in Ref.~\cite{Mitridate:2021ctr} which was previously limited to only the lowest energy, valence $\rightarrow$ conduction transitions. Lastly, in Sec.~\ref{subsec:annual_modulation} we demonstrate \textsf{EXCEED-DM}'s ability to compute modulation effects, and compute the annual modulation of the DM-electron scattering rate in the dark photon model.

\subsection{DM-Electron Scattering Rate Screened with Numeric Dielectric Function}
\label{subsec:dme_SI_numeric_dielectric}

Our first application is a new calculation of the DM-electron scattering rate in the canonical, kinetically mixed, dark photon model~\cite{Fabbrichesi:2020wbt}. As a quick review, this model contains a new, potentially broken, $U(1)$ symmetry whose gauge field is the dark photon, $A'_\mu$. The DM candidate is a new fermion, $\chi$, which is charged under this new $U(1)$ symmetry. The symmetries of the Lagrangian allow a kinetic mixing between the dark photon and photon,
\begin{align}
    \delta \mathcal{L} \supset \frac{\kappa}{2} F^{\mu \nu} F_{\mu \nu}' 
\end{align}
with coupling parameter $\kappa$. This interaction can be rotated to the mass basis by shifting the photon field, $A_\mu \rightarrow A_\mu + \kappa A'_\mu$, generating small ``dark electric" charges for the electrons,
\begin{align}
    \delta \mathcal{L} \supset -\kappa e A'_\mu \bar{e} \gamma^\mu e \, .
\end{align}

The DM-electron scattering rate can be shown to be in the form of Eq.~\eqref{eq:scatter_rate} (see Ref.~\cite{Griffin:2021znd} and App.~\ref{app:eff_L_to_F}) with,
\begin{align}
    \mathscr{F}_{IF} & = |\mathcal{T}_1|^2 \\
    f_\text{scr} & = \frac{1}{\varepsilon(q, \omega)} 
\end{align}
where we have assumed that the targets are isotropic enough to approximate $\hat{\mathbf{q}} \cdot \bm{\varepsilon} \cdot \hat{\mathbf{q}} \approx \varepsilon$, and $\mathcal{F}_\text{med}$ takes it standard form for the heavy/light mediator scenarios. 

While this rate has been computed for Si and Ge targets many times,~\cite{Graham:2012su, Lee:2015qva, Essig:2015cda, Catena:2021qsr, Griffin:2021znd, Catena2022, Griffin:2019mvc}, only recently were screening effects included numerically. The most recent calculation using \textsf{EXCEED-DM} uses an analytic model for a dielectric function. While the calculation in Refs.~\cite{Hochberg:2021pkt,Knapen:2021bwg} includes the screening effect numerically, the all-electron reconstruction effects~\cite{Griffin:2021znd, Liang:2018bdb}, which alter the electronic wave functions near the Fermi surface, were ignored, along with more deeply bound ``core" states, and high energy ``free" states, in the language of Ref.~\cite{Griffin:2021znd}. The calculation here includes these effects by computing the dielectric function with \textsf{EXCEED-DM}, with the same electronic configuration as used previously (which includes ``core" and ``free" states, in addition to the all-electron reconstruction corrected valence and conduction states).

The input for the dielectric function calculation is,
\begin{nice_box}{Input File}{BrickRed}
    [control] \\
        \begin{tabular}{ll}
            \inputtab calculation &= `dielectric' \\
            \inputtab
        \end{tabular}

    [numerics\_dielectric] \\
        \begin{tabular}{ll}
            \inputtab smear\_type &= `gauss' \\
            \inputtab widths &= 0.2, 0.1, 1000 \\
            \\
            \inputtab n\_E\_bins &= 1000 \\
            \inputtab n\_q\_bins &= 500 \\
            \inputtab E\_bin\_width &= 0.1 \quad \# eV \\
            \inputtab q\_bin\_width &= 0.1 \quad \# keV \\
        \end{tabular}
\end{nice_box}
where \textsf{smear\_type} is a specific broadening applied to the imaginary part of the Greens function, $G$, in Eq.~\eqref{eq:G_eq}, \textsf{widths} is a specific parameterization of the electron width (\textsf{widths} $= a, b, c$ corresponds to $\delta = \text{min}\left\{ a + b \omega, c \right\}$), and the other parameters define the binning parameters.
\begin{figure}[ht]
    \includegraphics[width=\textwidth]{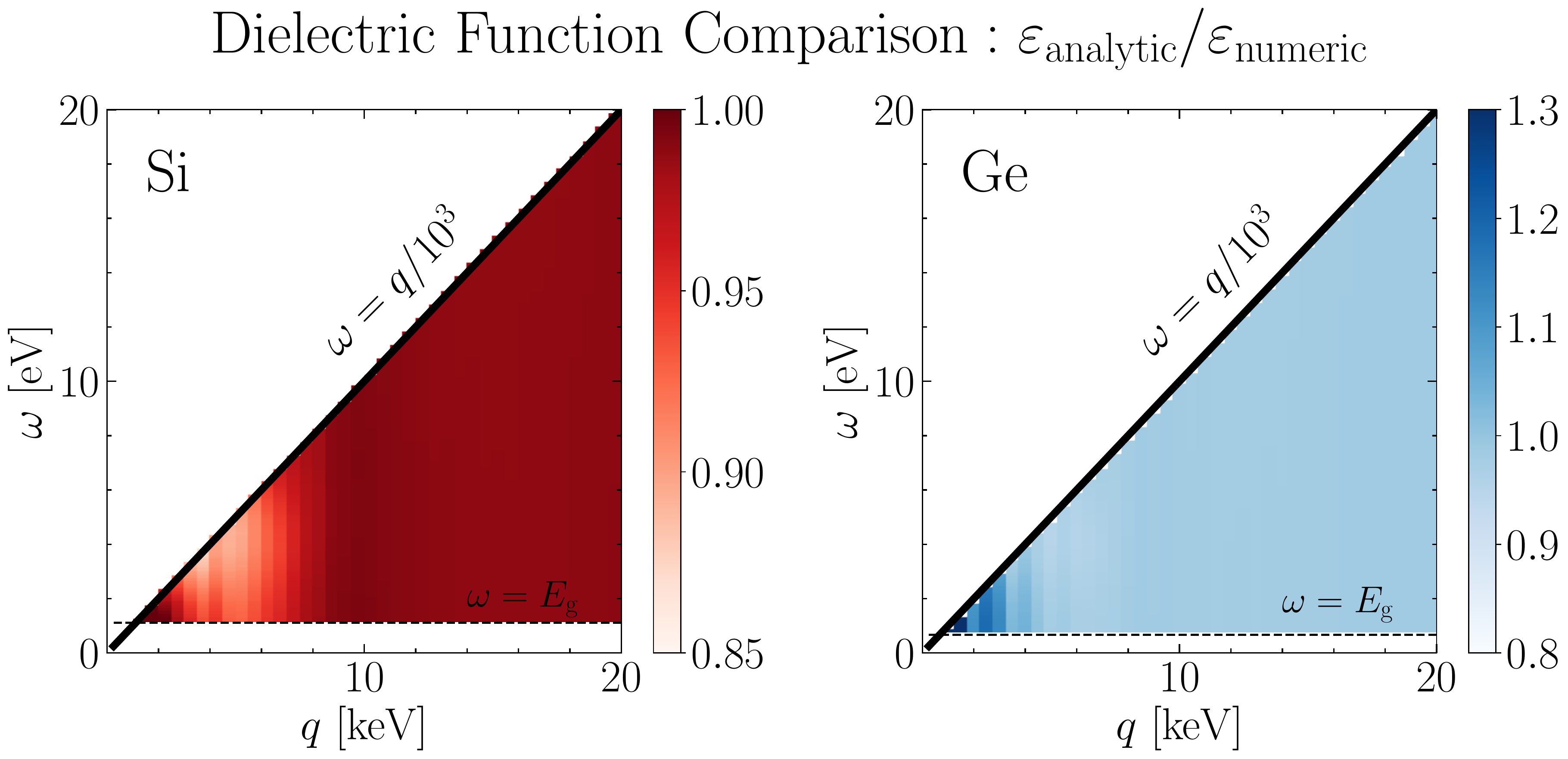}
    \caption{Comparison of the dielectric function computed analytically with the model from Ref.~\cite{Cappellini1993}, $\varepsilon_\text{analytic}$, to the dielectric function computed with \textsf{EXCEED-DM}, $\varepsilon_\text{numeric}$, for Si (left) and Ge (right) targets. These dielectric functions are used in the screening factor, Eq.~\eqref{eq:f_scr}, for the DM-electron scattering rate calculations. Since they are only used for scattering rate calculations, the only relevant differences between them are for kinematically allowed $q, \omega$, sectioned off by the black lines. The dashed black line corresponds to the band gap, $E_g$, and the solid black line bounds $\omega < q/v_\chi^\text{max}$, where $v_\chi^\text{max} \sim 10^{-3}$. There are $\mathcal{O}(10\%)$ differences at small $q, \omega$, which will propagate to slightly different DM-electron scattering rates for models that are screened. Both dielectric functions asymptote to one for large $q, \omega$.}
    \label{fig:dielectric_comparison}
\end{figure}
The resultant dielectric function is compared to the analytic model in Fig.~\ref{fig:dielectric_comparison},
\begin{align}
    \varepsilon(q, \omega) = 1 + \left[ \frac{1}{\epsilon_0 - 1} + \alpha \left( \frac{q}{q_{TF}} \right)^2  + \frac{q^2}{4 m_e^2 \omega_p^2} - \left( \frac{\omega}{\omega_p} \right)^2 \right]^{-1} \, , \label{eq:analytic_di}
\end{align}
defined in Ref.~\cite{Cappellini1993} with $\epsilon_0 = 11.3 \, (14), \alpha = 1.563, \omega = 16.6 \, (15.2)  \, \text{eV}, q_{TF} = 4.13 \, (3.99) \, \text{keV}$ for a Si (Ge) target. We find that the screening factor is dominated by low $q, \omega$ transitions, i.e., the dielectric function contribution from the ``core" and ``free" states is small, and in the kinematic regions where these transitions are dominant, the dielectric function can be approximated as one. Moreover, we only find $\mathcal{O}(10\%)$ changes between the analytic dielectric function model, and the numerically computed dielectric function. This will translate to an $\mathcal{O}(10\%)$ discrepancy in the DM-electron scattering rate when screened with the analytic model and \textsf{EXCEED-DM} computed dielectric function.

With the dielectric function computed, we now turn to computing the scattering rate with no screening, screening with the analytic dielectric function in Eq.~\eqref{eq:analytic_di}, and screening with the numerically computed dielectric function in Eq.~\eqref{eq:dielectric}. Additionally, we will study the dependence on a heavy and light mediator. The screening independent input groups are,
\begin{nice_box}{Input File}{BrickRed}
    [control] \\
        \begin{tabular}{ll}
            \inputtab calculation &= `binned\_scatter\_rate' \\
            \inputtab
        \end{tabular}

    [dm\_model] \\
        \begin{tabular}{ll}
            \inputtab FIF\_id &= 'SI' \\
            \inputtab mX\_logspace &= 20, 1e5, 1e10 \\
            \inputtab mX &= 1e6, 1e7, 1e8, 1e9 \\
            \inputtab med\_FF &= 0, 2 \\
            \inputtab
        \end{tabular}

    [astroph\_model] \\
        \begin{tabular}{ll}
            \inputtab v\_0\_km\_per\_sec &= 230 \\
            \inputtab v\_e\_km\_per\_sec &= 0, 0, 240 \\
            \inputtab v\_esc\_km\_per\_sec &= 600 \\
            \inputtab
        \end{tabular}

    [numerics\_binned\_scatter\_rate] \\
        \begin{tabular}{ll}
            \inputtab n\_q\_bins &= 1 \\
            \inputtab n\_E\_bins &= 4000 \\
            \inputtab q\_bin\_width &= 1   \# keV\\
            \inputtab E\_bin\_width &= 0.1 \# eV\\
        \end{tabular}
\end{nice_box}
where \textsf{mX} and \textsf{mX\_logspace} specify the DM masses to compute for, \textsf{med\_FF} are $- d \log \mathcal{F}_\text{med} / d \log q$ values, we have chosen the same velocity distribution parameters as previous studies~\cite{Griffin:2019mvc, Griffin:2021znd}, and we bin in $\omega$, since we will also study the binned scattering rate. The screening dependent groups are written,
\begin{nice_box}{Input File}{BrickRed}
    [screening] \\
        \begin{tabular}{ll}
            \inputtab type &= `analytic' \\
            \inputtab alpha &= 1.563 \\
            \inputtab e0 &= \textcolor{BrickRed}{11.3} \; \textcolor{NavyBlue}{14} \\
            \inputtab omega\_p &= \textcolor{BrickRed}{16.6} \; \textcolor{NavyBlue}{15.2} \\
            \inputtab q\_tf &= \textcolor{BrickRed}{4.13} \; \textcolor{NavyBlue}{3.99} \\
        \end{tabular}
\end{nice_box}
\begin{nice_box}{Input File}{BrickRed}
    [screening] \\
        \begin{tabular}{ll}
            \inputtab type &= `numeric' \\
            \inputtab dielectric\_filename &= \dots \\
            \inputtab E\_bin\_width &= 0.1 \# eV \\
            \inputtab q\_bin\_width &= 0.1 \# keV \\
            \inputtab width\_id &= 1
        \end{tabular}
\end{nice_box}
for the analytic screening, and numerically computed screening scenarios, respectively (no screening is the default, so no input groups have to be added). Entries in red are specific to Si targets, and those in blue specific to Ge targets. \textsf{dielectric\_filename} is the path to the previously computed dielectric function, \textsf{E\_bin\_width} and \textsf{q\_bin\_width} are the specific binning parameters used when computing the dielectric function (note that they match the \textsf{numerics\_dielectric} input previously discussed). \textsf{width\_id} specifies which width parameterization to use (from \textsf{widths} input for dielectric function calculation).

\begin{figure}[ht]
    \includegraphics[width=\textwidth]{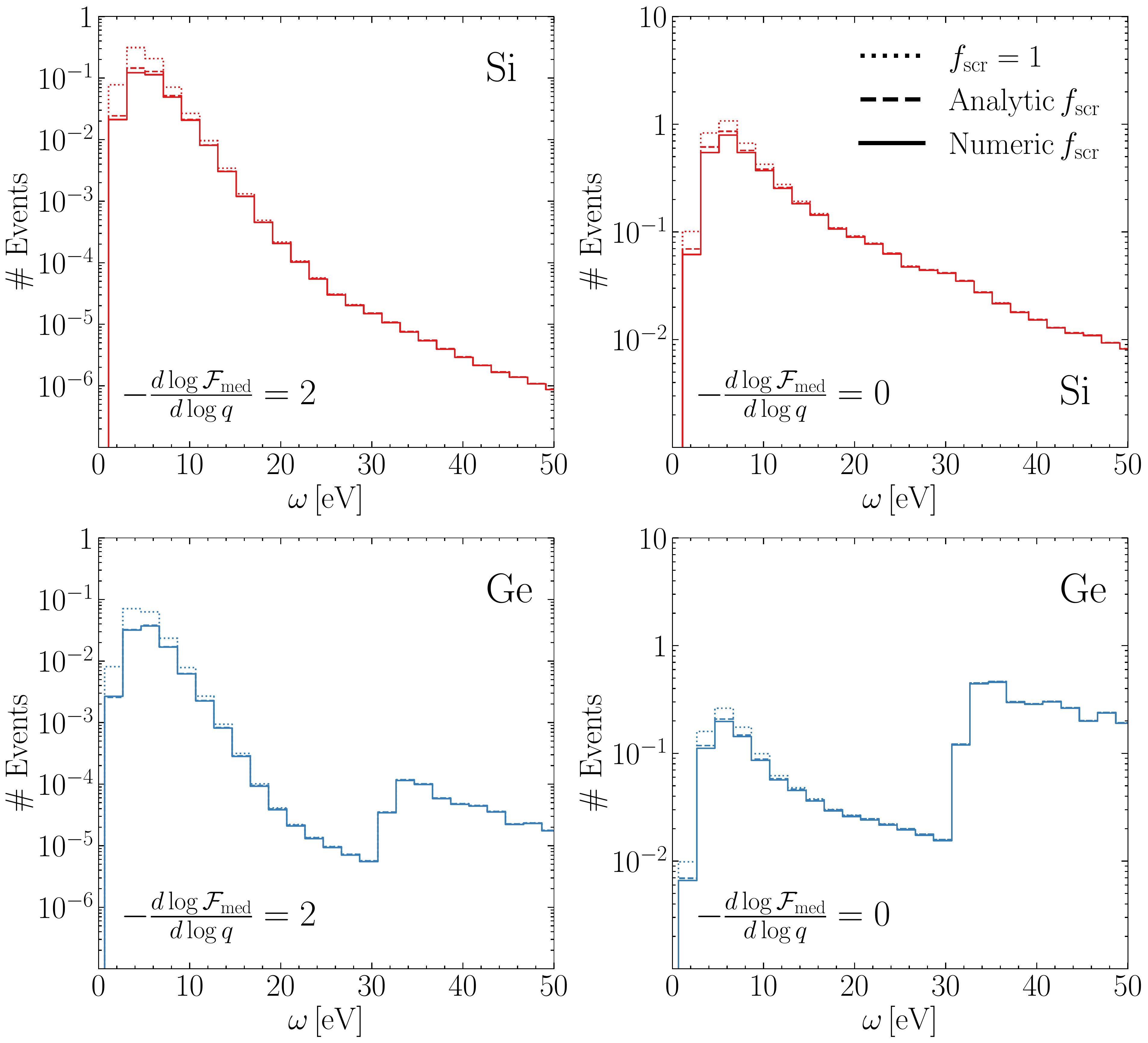}
    \caption{Prediction for the number of events in each energy bin, $\Delta \omega = 2 \; \text{eV}$, in Si (top row) and Ge targets (bottom row) for the light (left column) and heavy (right column) kinetically mixed dark photon model, assuming a kg-year exposure. The DM mass is $m_\chi = 1\,\text{GeV}$, and $\bar{\sigma}_e = 10^{-40} \, \text{cm}^2$. Dotted lines correspond to no screening, dashed lines correspond to screening with the analytic dielectric function in Eq.~\eqref{eq:analytic_di}, and solid lines correspond to screening with the numerically computed dielectric function. We see that while the inclusion of screening is important, as noted in Refs.~\cite{Hochberg:2021pkt,Knapen:2021bwg}, the difference between the analytic and numerically computed dielectric function, when used as the screening factor, is small.}
    \label{fig:screened_binned_rate_compare}
\end{figure}

\begin{figure}[ht]
    \includegraphics[width=\textwidth]{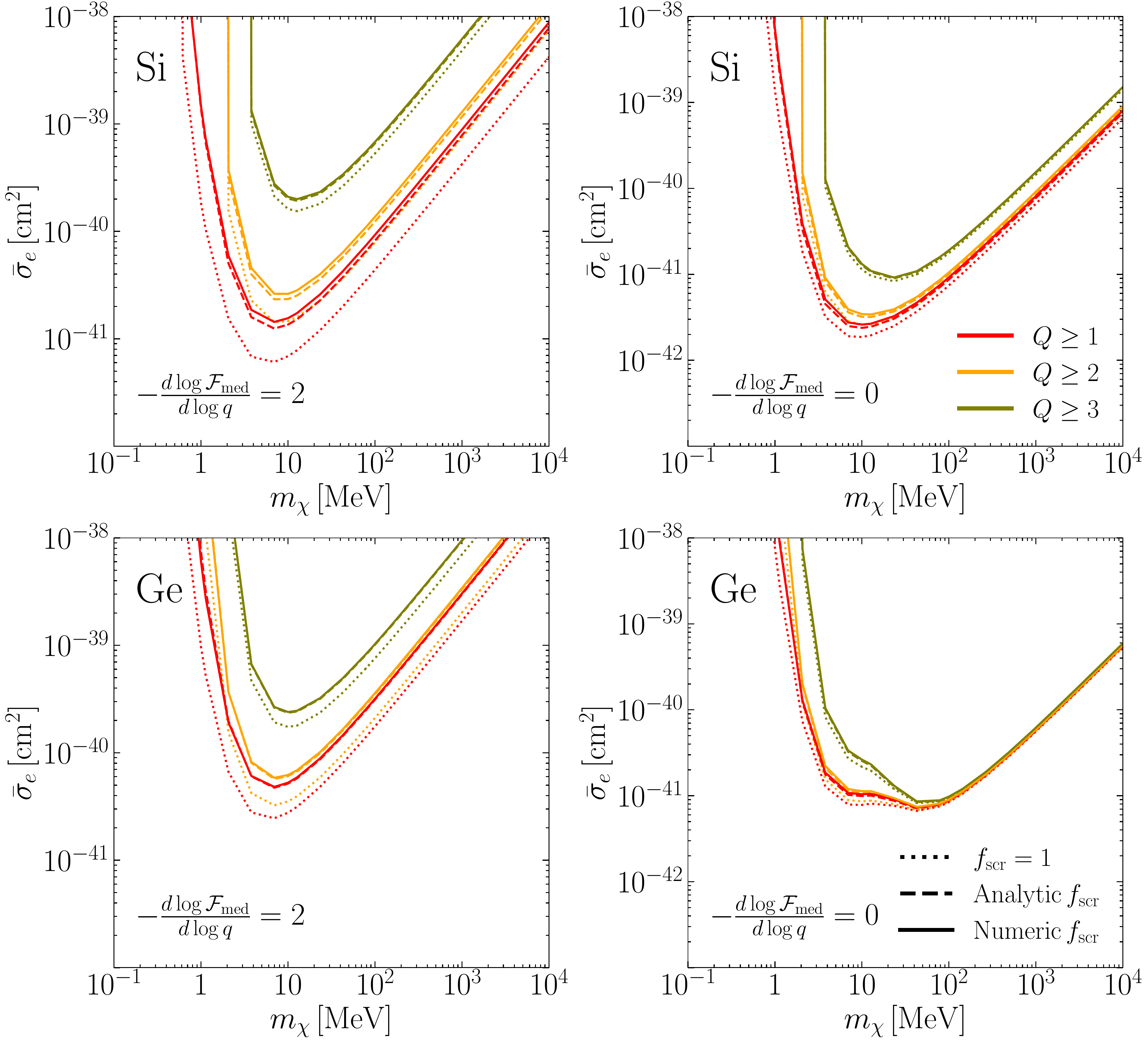}
    \caption{95\% C.L. constraints (3 events, no background) on DM-electron scattering cross section, $\bar{\sigma}_e$, in Si (top row) and Ge (bottom row) via kinetically mixed light (left column) and heavy (right column) dark photon models, assuming a kg-year exposure and different screening assumptions. Dotted lines correspond to no screening, as done in Ref.~\cite{Griffin:2019mvc}, dashed lines correspond to screening with the analytic model of the dielectric function, as done in Ref.~\cite{Griffin:2021znd}, and solid lines correspond to screening with the numerically computed dielectric function. The different colors correspond to different electron number thresholds, $Q \ge 1, 2, 3$ for the red, yellow, and green curves, respectively.}
    \label{fig:screened_cs_compare}
\end{figure}

With the inputs discussed, we now discuss the results of the DM-electron scattering rate calculation. In Fig.~\ref{fig:screened_binned_rate_compare} we compare the binned scattering rate in Si and Ge targets, assuming $m_\chi = 1\,\text{GeV}$, $\bar{\sigma}_e = 10^{-40} \, \text{cm}^2$, and different models for the screening factor, $f_\text{scr}$. We find that while screening is an important effect, especially for small $\omega$, the difference in using the analytic and numerically computed dielectric functions is small, $\mathcal{O}(10\%)$, with the difference being smaller in Ge than Si targets. In Fig.~\ref{fig:screened_cs_compare} we compare the cross section constraints and find similar results for the other DM masses considered. Additionally, we show that the effects of screening for different electron-hole pair number thresholds, $Q$, defined as,
\begin{align}
    Q = 1 + \left\lfloor \frac{\omega - E_g}{\varepsilon} \right\rfloor
\end{align}
where $E_g$ is the target band gap, taken to be $1.11 \, \text{eV}$ for Si and $0.67 \, \text{eV}$ for Ge, $\omega$ is the energy deposited in a single electron transition, and $\varepsilon$ is a material dependent quantity (not the dielectric function) which is taken to be $3.6 \; \text{eV}$ and $2.9 \; \text{eV}$ for Si and Ge targets respectively.

\subsection{DM-Electron Scattering Rate with Electron Velocity Dependent Operator}
\label{subsec:dme_VA}
\begin{figure}[ht]
    \includegraphics[width=\textwidth]{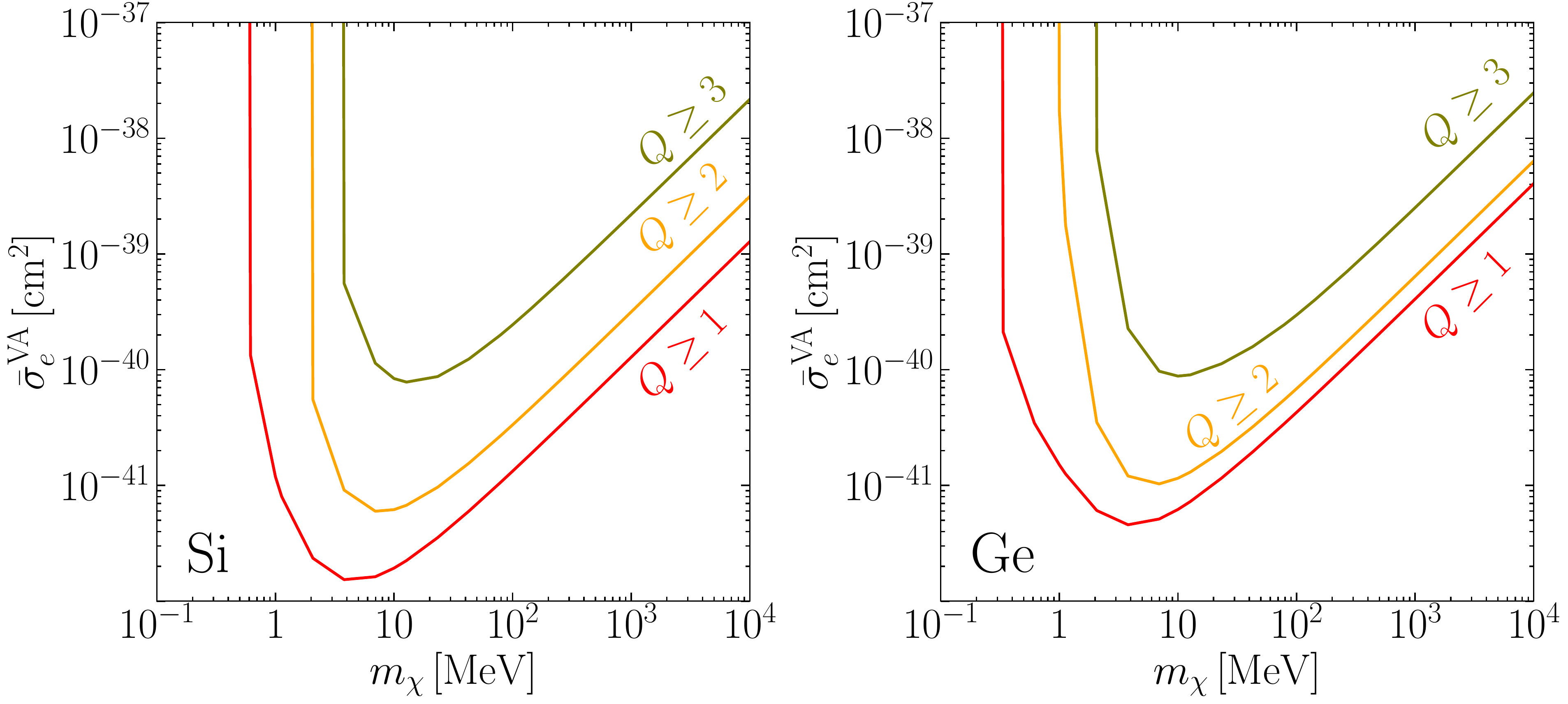}
    \caption{95\% C.L. (3 events, no background) constraints on the DM-electron scattering cross section, $\bar{\sigma}_e^{\text{VA}}$, given in terms of the couplings in Eq.~\eqref{eq:VA_cross_section}. We assume a kg-year exposure of Si (left) and Ge (right) targets, assuming the DM model given in Eq.~\eqref{eq:new_model}, and $V_\mu$ is light. The red, yellow, and green curves correspond to $Q \ge 1, 2, 3$, respectively.}
    \label{fig:VA_cs_reach}
\end{figure}

Our next calculation involves a new DM model, different than the kinetically mixed dark photon from Sec.~\ref{subsec:dme_SI_numeric_dielectric}. This DM model was chosen for two reasons: its simplicity, and the fact that the DM-electron scattering potential depends on the electron velocity. This generates a scattering rate dependent on the $\mathcal{T}_\mathbf{v}$ transition matrix element in Eq.~\eqref{eq:T_v}, and therefore only calculable with a complete spectrum of electronic states with \textsf{EXCEED-DMv1.0.0}.\footnote{While \textsf{QEDark-EFT}~\cite{Catena:2021qsr,Catena2022} also supports computing this DM-electron scattering rate, the calculation does not include all electron reconstructed valence and conduction states, and ignores the effects of `core' and `free' states, in the language of Ref.~\cite{Griffin:2021znd}. See Sec.~\ref{sec:introduction} for more details on the discrepancies between the codes.} The Lagrangian is given by,
\begin{align}
    \mathcal{L} \supset g_\chi V_\mu \bar{\chi} \gamma^\mu \chi + g_e V_\mu \bar{e} \gamma^\mu \gamma^5 e \, ,
    \label{eq:new_model}
\end{align}
where $V_\mu$ is a, light ($m_V \ll \text{keV}$), dark photon, and $\chi$ is the DM candidate. The scattering form factor, derived in Appendix~\ref{app:eff_L_to_F}, is given by,
\begin{align}
    \mathscr{F}_{IF} = \frac{1}{\alpha^2 m_e^2} \left( 4 m_e^2 |\mathcal{T}_\mathbf{v}|^2 + 2 m_e \mathcal{T}_1 (\mathbf{q} \cdot {\mathcal{T}_\mathbf{v}}^*) + 2 m_e {\mathcal{T}_1}^* (\mathbf{q} \cdot {\mathcal{T}_\mathbf{v}}) + q^2 |\mathcal{T}_1|^2  \right) \, .
    \label{eq:FIF_VA}
\end{align}
The reference cross section is,
\begin{align}
    \bar{\sigma}_e^\text{VA} \equiv \frac{\mu_{\chi e}^2}{4 \pi} \frac{\alpha^2 g_e^2 g_\chi^2}{\left( \alpha^2 m_e^2 + m_V^2 \right)^2} \, ,
    \label{eq:VA_cross_section}
\end{align}
and there is no screening, $f_\text{scr} = 1$. Fig.~\ref{fig:VA_cs_reach} contains the constraints on the reference cross section given in Eq.~\eqref{eq:VA_cross_section} in Si and Ge targets assuming a kg-year exposure. Explicit construction of a complete UV model, as well as further understanding of competing constraints, is beyond the scope of this work, see Ref.~\cite{Kahn:2016vjr} for a DM model building focused discussion.

The calculation for the DM-electron scattering rate in this model is identical to the unscreened input in Sec.~\ref{subsec:dme_SI_numeric_dielectric}, with \textsf{FIF\_id} changed to correspond to Eq.~\eqref{eq:FIF_VA}, and we focus on a light mediator only,
\begin{nice_box}{Input File}{BrickRed}
    [control] \\
        \begin{tabular}{ll}
            \inputtab calculation &= `binned\_scatter\_rate' \\
            \inputtab
        \end{tabular}

    [dm\_model] \\
        \begin{tabular}{ll}
            \inputtab FIF\_id &= 'VA1' \\
            \inputtab med\_FF &= 2 \\
            \inputtab
        \end{tabular}
\end{nice_box}

\subsection{Extended Absorption Calculation}
\label{subsec:extended_absorption}

Our next example concerns a different transition process, DM absorption~\cite{Chen2022,Derevianko:2010kz,Mitridate:2021ctr,Gelmini:2020xir,Hochberg:2017wce,Hochberg:2016sqx}. In these examples an incoming DM particle deposits its mass energy to the target, driving a, nearly, vertical transition. For some DM models, this absorption rate can be related to the measured, long wavelength dielectric function, $\varepsilon(0, \omega)$. Specifically, in isotropic targets, e.g., Si and Ge, the absorption rates for axion-like particle and vector DM, e.g., a dark photon, models can be related to $\varepsilon(0, \omega)$. However, even a simple model of scalar DM cannot be related, see Ref.~\cite{Mitridate:2021ctr} for more details. Therefore the first principle calculations provided by \textsf{EXCEED-DM} serves three purposes: 1) in DM models which \textit{cannot} be related to $\varepsilon(0, \omega)$, provide the only available calculation, 2) for DM models which \textit{can} be related, the calculation provides a data driven check on the input electronic configuration, and 3) for novel targets, which do not have a rigorously measured $\varepsilon(0, \omega)$, \textsf{EXCEED-DM} can facilitate a comparison with previously considered targets.

In this section we extend the constraints on DM absorption via electronic excitations to a larger mass range, up to $m_\phi = 1\,\text{keV}$, in the three DM models, scalar, pseudoscalar, and vector DM, considered in Ref.~\cite{Mitridate:2021ctr, Chen2022}. The interaction Lagrangians of each model are given in Table~\ref{tab:abs_rate_table}. The first principles calculations provided by \textsf{EXCEED-DM} for Si and Ge targets in Ref.~\cite{Mitridate:2021ctr} were limited to $m_\phi \lesssim 60 \; \text{eV}$ due to a more general implementation not being available in previous versions of \textsf{EXCEED-DM}. Here we use an expanded electronic configuration which includes deeply bound core states, and high energy free states, in the language of Ref.~\cite{Griffin:2021znd}.\footnote{There are minor differences in the electronic configurations used for the scattering rate calculation and the absorption rate calculations done here. This is due to the kinematics; because the transitions are vertical the states above and below the Fermi surface must be aligned on the same $\mathbf{k}$-grid in the 1BZ. Both electronic configurations are available here~\ZenodoElecConfigLink.}

The inputs for the calculation are,
\begin{nice_box}{Input File}{BrickRed}
    [control] \\
        \begin{tabular}{ll}
            \inputtab calculation &= `absorption\_rate' \\
            \inputtab
        \end{tabular}

    [dm\_model] \\
        \begin{tabular}{ll}
            \inputtab particle\_type &= `vector' \textcolor{NavyBlue}{(`ps', `scalar')} \\
            \inputtab mX\_logspace & = 100, 1, 1000
        \end{tabular}
    \\

    [numerics\_absorption\_rate] \\
        \begin{tabular}{ll}
            \inputtab smear\_type = `gauss' \\
            \inputtab widths = 0.2, 0.1, 1000
        \end{tabular}
\end{nice_box}
where the \textsf{particle\_type} input option specifies which DM model is considered in Table~\ref{tab:abs_rate_table}, and the entries in \textsf{numerics\_absorption\_rate} follow the same conventions as those for \textsf{numerics\_dielectric} in Sec.~\ref{subsec:dme_SI_numeric_dielectric}.

\begin{figure}[ht]
    \includegraphics[width=\textwidth]{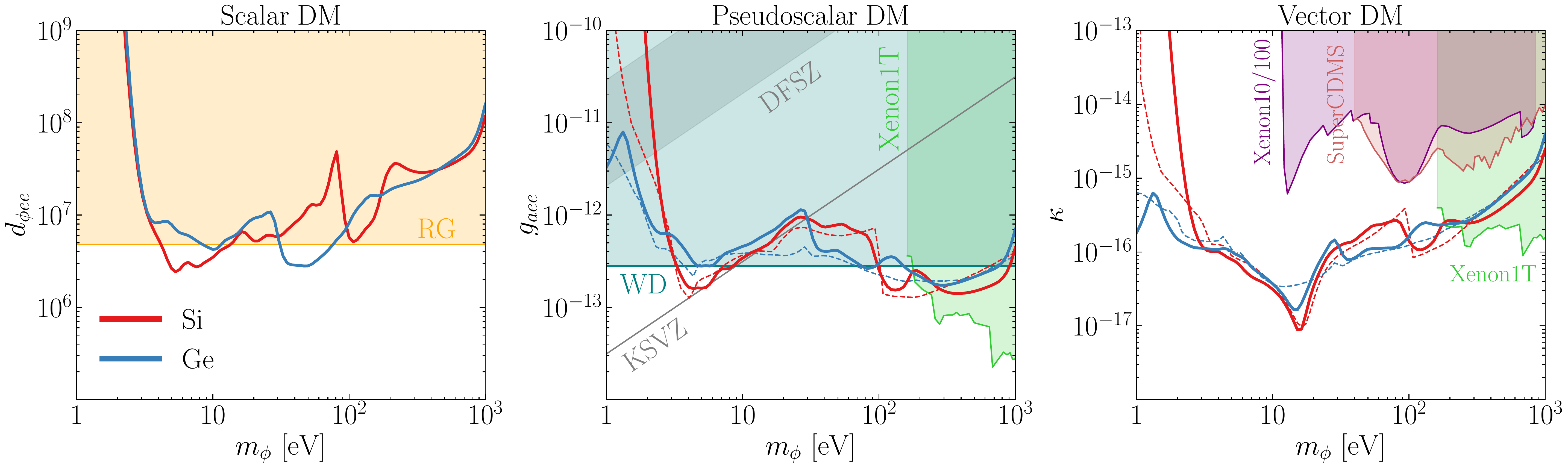}
    \caption{Projected 95\% C.L. constraints (3 events, no background) for scalar (left), pseudoscalar (middle), and vector (right) DM models with Si (red) and Ge (blue) targets with a kg-year exposure. Solid lines are constraints computed with \textsf{EXCEED-DM} and dashed lines are rescaled optical data from Ref.~\cite{1985}. The results for $m_\phi \lesssim 60 \, \text{eV}$ are in agreement with previous results from Ref.~\cite{Mitridate:2021ctr}, and constraints for $60 \, \text{eV} \lesssim m_\phi < 1\,\text{keV}$, which involve electronic states farther away from the Fermi surface, are the new addition to the calculation. Discrepancies with previous results at small $m_\phi$ are due to different smearing procedures. Agreement between the solid and dashed lines indicates a good approximation of the electronic configuration. RG (WD) stellar cooling bounds are taken from Ref.~\cite{Hardy:2016kme} (\cite{MillerBertolami:2014rka}). Direct detection constraints from Xenon10/100, Xenon1T, and SuperCDMS are taken from Refs.~\cite{Bloch:2016sjj},~\cite{XENON:2019gfn},~\cite{SuperCDMS:2019jxx} respectively. The gray lines in the middle panel are predictions of the DFSZ ($0.28 \leq \tan\beta \leq 140$) and KSVZ axion models~\cite{ParticleDataGroup:2020ssz}.}
    \label{fig:ext_absorption}
\end{figure}

The constraints on the couplings for the three DM models of interest, in Si and Ge targets assuming a kg-year exposure, are given in Fig.~\ref{fig:ext_absorption}. Note that the couplings have been conventionally redefined,
\begin{align}
    \kappa & = \frac{g_e}{e} \\
    g_{aee} & = g_e \\
    d_{\phi ee} & = \frac{M_\text{Pl}}{\sqrt{4 \pi} m_e} g_e \, ,
\end{align}
where $g_e$ is defined in the Lagrangian in Table~\ref{tab:abs_rate_table} and $M_\text{Pl}$ is the Planck mass. 

\subsection{Annual Modulation of DM-Electron Scattering Rate}
\label{subsec:annual_modulation}

In our last example, we will study a different aspect of DM-electron scattering in the kinetically mixed dark photon model than discussed in Sec.~\ref{subsec:dme_SI_numeric_dielectric}. Here we will study the annual modulation of the DM-electron scattering rate signal~\cite{Radick:2020qip,Essig:2015cda,Lee:2015qva,Drukier:1986tm,Green:2003yh,Freese:2012xd}. Annual modulation arises due to the Earths motion around the Sun; the Earths velocity relative to the DM wind changes as it revolves around the Sun, leading to changes in the DM velocity distribution. A signal which modulates annually would be a smoking gun signature of DM, since other backgrounds are not expected to have this feature.  

Annual modulation has been studied previously in Si and Ge targets~\cite{Radick:2020qip,Essig:2015cda,Lee:2015qva}, and the while the calculation here is an improvement due to more accurate modelling of the electronic structure in Si and Ge, the largest discrepancies are in DM models where these high momentum, high energy differences in the electronic structure models are important. Moreover, the calculation here uses the specific velocity distribution parameters which have been recommended to be used in direct detection calculations~\cite{Baxter:2021pqo}. One could also reproduce all the scattering rate results in Sec.~\ref{subsec:dme_SI_numeric_dielectric} with the output data~\ZenodoDataLink, and compare the dependence on the velocity distribution parameters. Our focus here will be on changing, $v_e$, while keeping $v_0, v_\text{esc}$ set to the recommended values~\cite{Baxter:2021pqo}. The input file is the same as in Sec.~\ref{subsec:dme_SI_numeric_dielectric}, with an updated \textsf{astroph\_model} group,
\begin{nice_box}{Input File}{BrickRed}
    [astroph\_model] \\
        \begin{tabular}{ll}
            \inputtab v\_0\_km\_per\_sec &= 238 \\
            \inputtab v\_esc\_km\_per\_sec &= 544 \\
            \inputtab v\_e\_km\_per\_sec &= 0, 0, 235 \\
            \inputtab v\_e\_km\_per\_sec &+= 0, 0, 250 \\
            \inputtab v\_e\_km\_per\_sec &+= 0, 0, 265
        \end{tabular}

\end{nice_box}
the $\mathbf{v}_e$ vectors we include in the calculation represent the average velocity, in the Galactic frame, $v_e = 250 \, \text{km}/\text{s}$, and $\pm 15 \, \text{km}/\text{s}$ fluctuations at its fastest/slowest over the year. Similar to Sec.~\ref{subsec:dme_SI_numeric_dielectric}, these calculations screen the DM-electron scattering rate with the numerically computed dielectric. The direction of the $\mathbf{v}_e$ vector is irrelevant for the isotropic targets of interest here, we choose the $\hat{\mathbf{z}}$ direction for simplicity. However, in anisotropic targets the direction will be important for daily modulation effects, and one can see how easy it is to add any list of $\mathbf{v}_e$ vectors to compute the daily modulation of the DM-electron scattering rate in such targets.

The main quantity of interest when studying annual modulation is the annual modulation fraction, $f_\text{mod}$. Let us define three rates, $R_+, R_-, R_0$ which are the rates when $v_e = v_e^0 + 15 \, \text{km/\text{s}}$, $v_e = v_e^0 - 15 \, \text{km}/\text{s}$, and $v_e = v_e^0$, where $v_e^0 = 250 \, \text{km}/\text{s}$, respectively. The annual modulation fraction is then defined as,
\begin{align}
    f_\text{mod} \equiv \frac{R_+ - R_-}{R_0} \, .
    \label{eq:f_mod}
\end{align}
\begin{figure}[ht]
    \includegraphics[width=\textwidth]{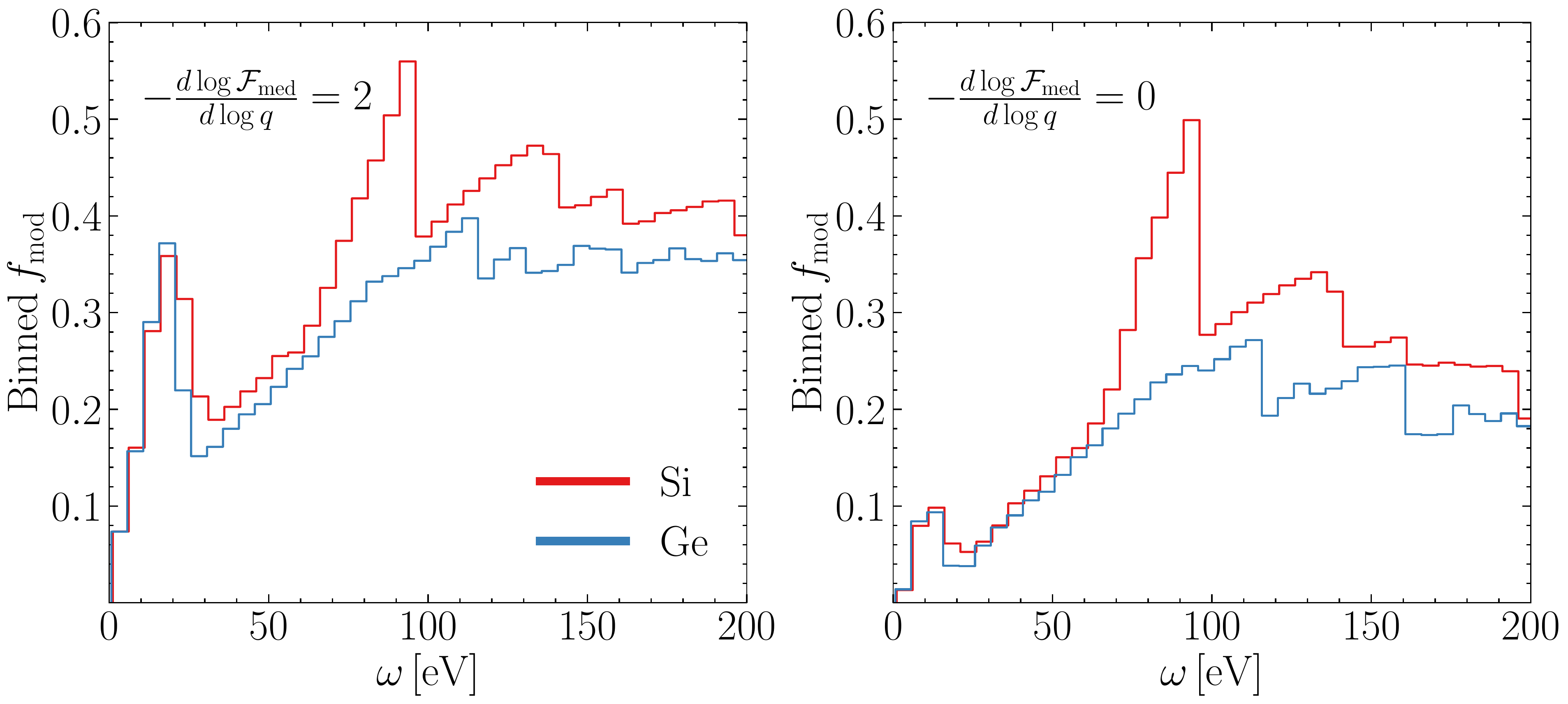}
    \caption{Annual modulation fraction, $f_\text{mod}$, in each energy bin, assuming $m_\chi = 1\,\text{GeV}$. In both Si (top row) and Ge (bottom row) targets, for both a light (left column) and heavy (right column) dark photon mediator, modulation fractions can be greater than $10 \%$, and stay relatively large in higher $\omega$ transitions.}
    \label{fig:ann_mod_binned_rate}
\end{figure}
\begin{figure}[ht]
    \includegraphics[width=\textwidth]{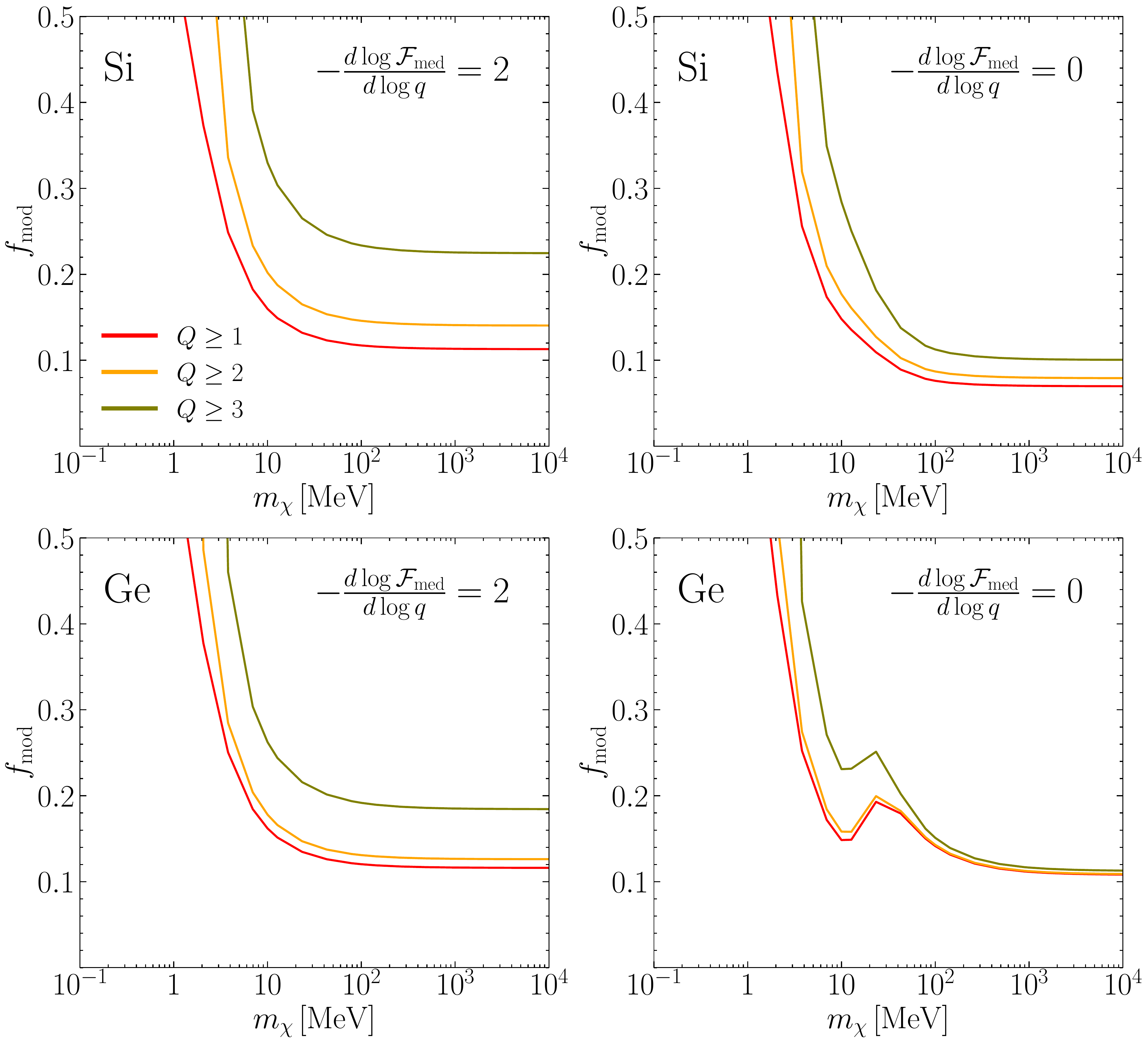}
    \caption{Annual modulation fraction, $f_\text{mod}$, given in Eq.~\eqref{eq:f_mod}, as a function of DM mass in Si (top row) and Ge (bottom row) targets. The different colors correspond to different $Q$ thresholds; the red, yellow, and green curves correspond to $Q \ge 1, 2, 3$, respectively.}
    \label{fig:ann_mod_fmod}
\end{figure}
In Fig.~\ref{fig:ann_mod_binned_rate} we study $f_\text{mod}$, binned in energy deposition, for $m_\chi =1\, \text{GeV}$ for Si and Ge targets. We find that the annual modulation fraction in both targets is generally greater than $10\%$. While the overall size of the modulation is consistent with previous studies~\cite{Lee:2015qva, Essig:2015cda}. Interestingly, the behavior extends to larger energy depositions, meaning that high threshold experiments will see the same amount of annual modulation as low threshold experiments (albeit at lower overall event rate).

In Fig.~\ref{fig:ann_mod_fmod} we study the annual modulation along a different axis; specifically, $f_\text{mod}$ versus DM mass for different electron-hole pair thresholds, $Q$. For low DM masses $f_\text{mod}$ becomes asymptotically large which is simply due to the shift in kinematically allowed DM masses. The lightest DM mass which can scatter is given by $m_\chi \gtrsim 2 E_g / v_\text{max}^2$, where $v_{\text{max}} = v_e + v_\text{esc}$. Therefore for different values of $v_e$, the lowest detectable DM mass changes. Higher values of $v_e$ correspond to smaller DM masses, and therefore $R_0$ becomes kinematically inaccessible before $R_+$ as the DM mass decreases. This causes $f_\text{mod}$ to diverge. At higher masses, $f_\text{mod}$ generally levels off, except when deeply bound states get involved, e.g., the 3d shell in Ge, whose effects can be seen at $m_\chi \approx 20 \, \text{MeV} $ in the bottom right panel of Fig.~\ref{fig:ann_mod_fmod}. The dependence on $Q$ is also easily understood from Fig.~\ref{fig:ann_mod_binned_rate}; increasing the threshold picks out the pieces of the binned scattering rate which have larger $f_\text{mod}$. 

\section{Summary and Future Development}
\label{sec:summary}

\textsf{EXCEED-DM} has already been shown to be useful for a wide range of DM-electron interaction calculations: from target comparison studies~\cite{Trickle:2019nya} of DM-electron scattering, proof-of-principle daily modulation signals in anisotropic targets~\cite{Trickle:2019nya}, extending the DM-electron scattering rate calculation to include higher energy and momentum contributions~\cite{Griffin:2021znd}, performing first principles DM-electron absorption rate calculations~\cite{Mitridate:2021ctr}, and DM-electron interaction rate calculations in spin orbit coupled target~\cite{Chen2022}. More recently the CDEX collaboration has used it to place constraints with their Ge detector~\cite{Zhang2022}, and \textsf{EXCEED-DM} has also been used in studies of non-standard neutrino interactions~\cite{Dutta2022}. Clearly, \textsf{EXCEED-DM} has proved to be a useful tool for studying new physics, even before the \textsf{v1.0.0} release. 

The purpose of releasing a \textsf{v1.0.0}, and this manuscript, is to
\begin{itemize}
    \item Introduce users to the inputs and formalism \textsf{EXCEED-DMv1.0.0} works with.
    \item Rewrite the \textsf{EXCEED-DMv0.3.0} codebase to provide a split between the electronic state approximations and observables being computed. This allows new calculations to be added without rewriting the transition matrix elements from scratch for each electronic state approximation.
    \item Perform new calculations which have been implemented, since \textsf{v0.3.0}, by complementing previous results. These new calculations represent the best available projections for Si and Ge based detectors used in SuperCDMS~\cite{SuperCDMS:2020ymb,SuperCDMS:2018mne,CDMS:2009fba}, DAMIC~\cite{DAMIC:2020cut,DAMIC:2016qck,Settimo:2020cbq,DAMIC:2015znm,DAMIC:2019dcn}, EDELWEISS~\cite{EDELWEISS:2018tde,EDELWEISS:2019vjv,EDELWEISS:2020fxc}, SENSEI~\cite{Crisler:2018gci,SENSEI:2020dpa,SENSEI:2019ibb}, and CDEX~\cite{Zhang2022}.
\end{itemize}
In Sec.~\ref{subsec:dme_SI_numeric_dielectric} we compute the dielectric function with the same electronic configuration we use to compute the scattering rate in Ref.~\cite{Griffin:2021znd}, providing the most accurate calculation of the DM-electron scattering rate in the kinetically mixed dark photon model. Then in Sec.~\ref{subsec:dme_VA} we compute the DM-electron scattering rate from an interaction that depends on the electron velocity; a model that was not calculable previously with a complete electronic configuration. In Sec.~\ref{subsec:extended_absorption} we extend the results of the DM-electron absorption calculation to higher DM masses, up to $m_\phi = 1\,\text{keV}$, and find good agreement with the measured dielectric function when expected. Lastly, in Sec.~\ref{subsec:annual_modulation} we study the annual modulation of the DM-electron scattering rate in the kinetically mixed dark photon model, and confirm previous findings~\cite{Lee:2015qva, Essig:2015cda} of $\mathcal{O}(10\%)$ annual modulation fractions. We also find that they persist for higher thresholds. We make public not only the \textsf{EXCEED-DM} codebase~\GithubLink, but also the electronic configurations used~\ZenodoElecConfigLink and the output of all the calculations performed here~\ZenodoDataLink so that results here can be reproduced, and used in future work. Additionally, we have created a documentation website~\DocLink full of specific details about all inputs and outputs.

Future development of \textsf{EXCEED-DM} will focus on generalizing these results. From the electronic configuration aspect, adding a broader set of electronic wave function approximations to more accurately describe a variety of targets. Adding more calculations for different observables, a wider range of DM models, as well as velocity distributions. From the experimental side, connecting the \textsf{EXCEED-DM} output to more experimental projections, including backgrounds. Adding these components will allow experimentalists to maximize their search over DM model space, and in the event of a signal, aid in maximally understanding the fundamental properties of DM.

\acknowledgments
I would like to thank Andrea Mitridate, Zhengkang Zhang and Kathryn M. Zurek for their help developing previous versions of \textsf{EXCEED-DM}, Sin\'ead Griffin and Katherine Inzani for their expertise in DFT calculations, and Bogdan Dobrescu for helpful discussions. 
This manuscript has been authored by Fermi Research Alliance, LLC under Contract No. DE-AC02-07CH11359 with the U.S. Department of Energy, Office of Science, Office of High Energy Physics. Early development of \textsf{EXCEED-DM} was supported by the Quantum Information Science Enabled Discovery (QuantISED) for High Energy Physics (KA2401032).
The computations presented here were conducted on a combination of the Caltech High Performance Cluster, partially supported by a grant from the Gordon and Betty Moore Foundation, and the Wilson-Institutional Cluster at Fermilab. 

\appendix

\section{Installation of EXCEED-DM}
\label{sec:installation}

\textsf{EXCEED-DM} is hosted on Github~\GithubLink, and releases are stored in \textsf{EXCEED-DM.tar.gz} files.To install \textsf{EXCEED-DM}, the \textsf{.tar.gz} file must first be downloaded and extracted. There are a few ways to do this. To download the latest release from the command line you can use \textsf{curl},\footnote{Specific versions can be downloaded in a similar way, with slightly different command. See the documentation~\DocLink for more details.}
\begin{nice_box}{Command Line}{ForestGreen}
    curl -LO https://github.com/tanner-trickle/EXCEED-DM/releases/latest/download/EXCEED-DM.tar.gz
\end{nice_box}
Alternatively, you can visit the \textsf{Releases} section on the Github page, \href{https://github.com/tanner-trickle/EXCEED-DM/releases}{here}, and download a specific release. Once the package has been downloaded, simply extract it to a folder titled, e.g., \textsf{EXCEED-DM}. Assuming you are in the folder with the \textsf{EXCEED-DM.tar.gz} file,
\begin{nice_box}{Command Line}{ForestGreen}
mkdir EXCEED-DM \&\& cd EXCEED-DM \\
tar -xzvf ../EXCEED-DM.tar.gz
\end{nice_box}

With \textsf{EXCEED-DM} downloaded and extracted, we now need to compile the main \textsf{exdm} program. Before compiling make sure all of the necessary prerequisite software is installed: \textsf{BLAS}, \textsf{LAPACK}, \textsf{CMake}, \textsf{FFTW3}, \textsf{HDF5}, \textsf{OpenMPI}, along with a \textsf{Fortran} compiler. Once the prerequisites are installed the main executable, \textsf{exdm}, can be compiled with
\begin{nice_box}{Command Line}{ForestGreen}
mkdir build \&\& cd build \\
cmake .. \\
make
\end{nice_box}
assuming you start in the folder you extracted the \textsf{EXCEED-DM.tar.gz} file to. To check that \textsf{exdm} was built correctly we can try to run it with the,
\begin{nice_box}{Command Line}{ForestGreen}
./exdm
\end{nice_box}
command, which should return something similar to
\begin{nice_box}{Console Output}{ForestGreen}
\begin{verbnobox}[\sffamily]
--------------------------------------------------------------------------------

    EXCEED-DM - v1.0.0

    Running on 1 processors
    Compiled with GCC version 11.1.0

    Started at 16:52:19.532 10/24/2022

--------------------------------------------------------------------------------

No input file specified, aborting.
\end{verbnobox}
\end{nice_box}
For further installation instructions see the documentation~\DocLink.

\section{Effective Lagrangian to Scattering Form Factor}
\label{app:eff_L_to_F}

\textsf{EXCEED-DM} can compute DM-electron scattering rates that can be written in terms of the transition matrix elements in Eqs.~\eqref{eq:T_1} -~\eqref{eq:T_vxs}. The main theoretical problem is then going from a UV Lagrangian to the scattering form factor, $\mathscr{F}_{IF}$, used in Eq.~\eqref{eq:scatter_rate}. At a high level, this is done in two steps: first, find the non-relativistic limit (NR) of the UV Lagrangian (see Refs.~\cite{Mitridate:2021ctr, Catena2022} for a more detailed discussion of this step). Second, from the NR effective Lagrangian, use Fermi's Golden rule to compute the scattering rate, and reference cross section, and then manipulate the resulting formula to the form used in Eq.~\eqref{eq:scatter_rate}. The first step is well outside the scope of this paper. In this appendix we will focus on the second step, starting from a fairly generic NR effective Lagrangian, and deriving the scattering form factor directly in terms of the operators appearing in the NR effective Lagrangian. To illustrate the usefulness of this approach we will derive the scattering form factors for the two models focused on in the main text: the kinetically mixed dark photon model, Secs.~\ref{subsec:dme_SI_numeric_dielectric},~\ref{subsec:annual_modulation}, and the `V-A' DM model in Sec.~\ref{subsec:dme_VA}.

Our starting point is the NR effective Lagrangian of a DM fermion, $\chi$, coupling to the electron, $\psi$, 
\begin{align}
    \mathcal{L}^\text{NR}_\text{eff} \supset \chi^\dagger \psi^\dagger \mathbf{\mathscr{O}} \chi \psi
    \label{eq:L_NR}
\end{align}
where $\psi, \chi$ are the two component NR electron and DM fields, respectively, and $\mathscr{O}$ is the interaction operator. $\chi$ and $\psi$ each have units of $\text{eV}^{3/2}$, and $\mathscr{O}$ has units of $\text{eV}^{-2}$. Assume an incoming DM particle with momentum $\mathbf{p}$ and spin $s$ scatters off an electronic state $| I \rangle$, creating to an outgoing DM particle with momentum, $\mathbf{p}' = \mathbf{p} - \mathbf{q}$ ($q$ is the momentum transferred to the target) and a final electronic state $| F \rangle$. Fermi's Golden Rule tells us that the transition rate due to the Lagrangian in Eq.~\eqref{eq:L_NR} is given by,
\begin{align}
    \Gamma_{I, \mathbf{p}, s \rightarrow F, s'} = \frac{2 \pi}{V} \int \frac{d^3\mathbf{q}}{(2 \pi)^3} \delta(E_F - E_I - \omega_\mathbf{q}) | \langle F | e^{i \mathbf{q} \cdot \mathbf{x} }\mathscr{O}_{ss'} | I \rangle |^2 \, ,
    \label{eq:Gamma_1}
\end{align}
where $E_I$, $E_F$ are the energies of the electronic states, $\omega_\mathbf{q} = \mathbf{q} \cdot \mathbf{v} - \frac{q^2}{2 m_\chi}$, $\mathbf{v}$ is the incoming DM velocity, and $m_\chi$ is the DM mass. The volume, $V$, and exponential factor come from the incoming and outgoing DM states, $\chi | \mathbf{p}, s \rangle = V^{-1/2} e^{i \mathbf{p} \cdot \mathbf{x}} \xi_s$, and $\xi_\uparrow = \begin{pmatrix}1 \\ 0\end{pmatrix}$, $\xi_\downarrow = \begin{pmatrix} 0 \\ 1 \end{pmatrix}$. $\mathscr{O}_{ss'}$ is simply the $s,s'$ component of $\mathcal{O}$ in Eq.~\eqref{eq:L_NR}, i.e., $\mathscr{O}_{ss'} = \xi_{s'}^\dagger \mathscr{O} \xi_s$.

Assuming the DM is unpolarized we spin average (sum) over the initial (final) DM spin states in Eq.~\eqref{eq:Gamma_1} to get the spin average interaction rate,
\begin{align}
    \Gamma_{I, \mathbf{p} \rightarrow F} = \frac{2\pi}{V} \int \frac{d^3\mathbf{q}}{(2 \pi)^3} \delta(E_F - E_I - \omega_\mathbf{q}) \overline{| \langle F | e^{i \mathbf{q} \cdot \mathbf{x}} \mathscr{O}_{ss'} | I \rangle |^2} \, ,
\end{align}
where the overline notation indicates $(1/2) \sum_{ss'}$. Assuming the DM velocity distribution is given by $f_\chi(\mathbf{v})$ and $g(\mathbf{q}, \omega) \equiv 2 \pi \int d^3 \mathbf{v} f_\chi(\mathbf{v}) \delta(\omega - \omega_\mathbf{q})$, the average transition rate between two electronic states becomes,
\begin{align}
    \Gamma_{I \rightarrow F} = \frac{1}{V} \int \frac{d^3\mathbf{q}}{(2 \pi)^3} g(\mathbf{q}, E_F - E_I) \overline{| \langle F | e^{i \mathbf{q} \cdot \mathbf{x}} \mathscr{O}_{ss'} | I \rangle |^2} \, .
\end{align}
The transition rate per kg-year, $R$, is then simply this interaction rate multiplied by the number of DM particles in the target, divided by the target mass, and summed over initial and final states,
\begin{align}
    R = \frac{\rho_\chi}{m_\chi \rho_T V} \sum_{IF} \int \frac{d^3\mathbf{q}}{(2 \pi)^3} g(\mathbf{q}, E_F - E_I) \overline{| \langle F | e^{i \mathbf{q} \cdot \mathbf{x}} \mathscr{O}_{ss'} | I \rangle |^2} \, ,
    \label{eq:app_R}
\end{align}
where $\rho_\chi$ is the DM density, and $\rho_T$ is the target density. Eq.~\eqref{eq:app_R} is most useful if one is interested in finding the rate as a function of the Lagrangian parameters in the NR EFT since it is \textit{directly} related to the $\mathscr{O}$ appearing in Eq.~\eqref{eq:L_NR}. However, it is useful, and common practice, to write this rate in terms of a reference cross section, which is simply a function of the underlying Lagrangian parameters. In terms of $\mathscr{O}$, the reference cross section is given by,
\begin{align}
    \bar{\sigma}_e \equiv \frac{\mu_{\chi e}^2}{4 \pi} \sum_{ss', \sigma \sigma'} |\mathscr{O}_{ss',\sigma\sigma'}|^2 \, ,
\end{align}
where $\mu_{\chi e}$ is the DM, electron reduced mass. This function is evaluated at reference momentum scales, e.g., $q = q_0 = \alpha m_e$, and ignores screening effects, $\varepsilon = 1$. We can now write Eq.~\eqref{eq:app_R} in terms of the reference cross section. At the risk of abusing notation, we define,
\begin{align}
    \hat{\mathscr{O}}_{ss'} = \frac{\mathscr{O}_{ss'}}{\sqrt{ \frac{1}{4} \sum_{ss', \sigma \sigma'} |\mathscr{O}_{ss'\sigma\sigma'}|^2}} \, ,
\end{align}
and the rate can be written as,
\begin{align}
    R = \frac{\pi \rho_\chi \bar{\sigma}_e}{\mu_{\chi e}^2 m_\chi \rho_T V} \sum_{IF} \int \frac{d^3\mathbf{q}}{(2 \pi)^3} g(\mathbf{q}, E_F - E_I) \overline{| \langle F | e^{i \mathbf{q} \cdot \mathbf{x}} \hat{\mathscr{O}}_{ss'} | I \rangle |^2} \, .
\end{align}
From here, the connection to Eq.~\eqref{eq:scatter_rate} is a simple rearrangement, and we see that the scattering form factor is simply $\overline{| \langle F | e^{i \mathbf{q} \cdot \mathbf{x}} \hat{\mathscr{O}}_{ss'} | I \rangle |^2}$ with the propagator dependence factored out,
\begin{align}
    \mathscr{F}_{IF} = \frac{1}{f_\text{scr}^2 \mathcal{F}_\text{med}^2} \overline{| \langle F | e^{i \mathbf{q} \cdot \mathbf{x}} \hat{\mathscr{O}}_{ss'} | I \rangle |^2} \, .
    \label{eq:scatter_form_factor_derivation}
\end{align}

With Eq.~\eqref{eq:scatter_form_factor_derivation} we can compute the scattering form factor for the dark photon and `VA' DM model. We begin with the dark photon model which has the UV Lagrangian,
\begin{align}
    \mathcal{L} \supset g_\chi V_\mu \bar{\chi} \gamma^\mu \chi + g_e V_\mu \bar{e} \gamma^\mu e \, ,
\end{align}
where $V_\mu$ is the dark photon and it has standard kinetic terms. If the coupling $g_e$ is due to kinetic mixing, then $g_e = e \kappa$, where $\kappa$ is the kinetic mixing parameter. It can be shown that the NR EFT Lagrangian and $\mathscr{O}$, are,
\begin{align}
    \mathcal{L}_\text{eff}^\text{NR} & \supset \frac{g_e g_\chi}{\varepsilon(q, \omega) \left(q^2 + m_{V}^2 \right)} \chi^\dagger \psi^\dagger \psi \chi \\
    \mathscr{O}_{ss', \sigma\sigma'} & = \frac{g_e g_\chi}{\varepsilon(q, \omega) \left( q^2 + m_{V}^2 \right)} \delta_{ss'} \delta_{\sigma\sigma'} \, ,
\end{align}
for a material with an isotropic dielectric function, where $m_V$ is the dark photon mass, and $\omega$ is the energy transferred to the target. We will find the scattering form factor in two limits, for a heavy and light dark photon. We begin by computing $\hat{\mathscr{O}}_{ss'}$,
\begin{align}
    \hat{\mathscr{O}}_{ss'} = \delta_{ss'} \frac{1}{\varepsilon(q, \omega)} \begin{cases}  \frac{q_0^2}{q^2} & \text{light} \\
                                            1 & \text{heavy} \end{cases} \, .
\end{align}
In both cases, $f_\text{scr} = 1/\varepsilon$, and $\mathcal{F}_\text{med}$ takes its standard form for heavy and light mediators. The scattering form factor is the same in both the light and heavy dark photon models,
\begin{align}
    \mathscr{F}_{IF} = |\mathcal{T}_1|^2 \, .
\end{align}

Moving on to the VA DM model we follow the same steps, first by writing down the UV Lagrangian,
\begin{align}
    \mathcal{L} \supset g_\chi V_\mu \bar{\chi} \gamma^\mu \chi + g_e V_\mu \bar{e} \gamma^\mu \gamma^5 e \, .
\end{align}
Following Ref.~\cite{Mitridate:2021ctr, Trickle:2019nya} after taking the NR limit of the currents and integrating out the dark photon, the NR effective Lagrangian is,
\begin{align}
    \mathcal{L}_\text{eff}^\text{NR} \supset \frac{g_\chi g_e}{q^2 + m_V^2} \, \chi^\dagger \psi^\dagger \left[ \frac{\mathbf{K}_e\cdot \bm{\sigma}_e}{2 m_e}  - \frac{\mathbf{K}_\chi\cdot \bm{\sigma}_e}{2m_\chi} + \frac{i \bm{\sigma}_e \cdot \left( \mathbf{q} \times \bm{\sigma}_\chi \right)}{2m_\chi} \right] \psi \chi \, ,
    \label{eq:VA_model_eff_L}
\end{align}
where $\mathbf{K}_e = 2 \mathbf{k}_e + \mathbf{q}$, $\mathbf{k}_e$ is the electron momentum, $\mathbf{K}_\chi = 2\mathbf{k}_\chi - \mathbf{q}$, $\mathbf{k}_\chi$ is the DM momentum, and $\sigma_\chi$ ($\sigma_e$) are the Pauli matrices acting in DM (electron) space. Contrary to the kinetically mixed dark photon model, there is no screening. Since the dark photon couples to electrons via the spin operator, $\bm{\sigma}_e$, there is no mixing between the dark photon-photon to generate screening terms. 

The NR effective Lagrangian in Eq.~\eqref{eq:VA_model_eff_L} can be further simplified since the last two terms are of order $\mathcal{O}(v_\chi)$, whereas the first term is $\mathcal{O}(v_e)$. Since $v_e \sim \alpha \sim 10^{-2}$ the first term will dominate and we will approximate the interaction with just the first term. Therefore,
\begin{align}
    \hat{\mathscr{O}}_{ss'} = \mathcal{F}_\text{med} \frac{\mathbf{K}_e \cdot \bm{\sigma}_e}{K_e^0} \delta_{ss'}
\end{align}
where $K_e^0$ is some reference momentum, which we take to be $\alpha m_e$. Therefore the scattering form factor is,
\begin{align}
    \mathscr{F}_{IF} = \frac{1}{(\alpha m_e)^2} \left[ 4 m_e^2 \left| \mathcal{T}_{\mathbf{v} \cdot \bm{\sigma}} \right|^2 + 2 m_e \mathcal{T}_{\mathbf{v} \cdot \bm{\sigma}} \left( \mathbf{q} \cdot \mathcal{T}_{\bm{\sigma}} \right)^* + 2 m_e \mathcal{T}_{\mathbf{v} \cdot \bm{\sigma}}^* \left( \mathbf{q} \cdot \mathcal{T}_{\bm{\sigma}} \right) + | \mathbf{q} \cdot \mathcal{T}_{\bm{\sigma}} |^2 \right] \, .
    \label{eq:FIF_VA_model_0}
\end{align}
This scattering form factor can be further simplified for targets which have spin degenerate eigenstates, since the electron spin sum can be performed analytically. Performing the electron spin sum, $(1/2) \sum_{\sigma \sigma'}$, gives,
\begin{align}
    \mathscr{F}_{IF} = \frac{1}{(\alpha m_e)^2} \left[ 4 m_e^2 \left| \mathcal{T}_{\mathbf{v}} \right|^2 + 2 m_e \left( \mathbf{q} \cdot \mathcal{T}_{\mathbf{v}} \right) \mathcal{T}_1^* + 2 m_e \left( \mathbf{q} \cdot \mathcal{T}_{\mathbf{v}}^* \right) \mathcal{T}_1 + q^2 | \mathcal{T}_1|^2 \right] \, .
    \label{eq:FIF_VA_model}
\end{align}
where $I, F$ are now just the band labels. 

\section{Transition Matrix Elements for Bloch States}
\label{app:T_for_bloch}

We will discuss the explicit derivation of the transition matrix elements, $\langle I | \mathcal{O} | F \rangle$ for the Bloch states discussed in Sec.~\ref{subsec:elec_state_config}. Note that since \textsf{bloch\_PW\_basis}, \textsf{bloch\_STO\_basis}, \textsf{bloch\_single\_PW} are simply different bases of the same approximation, these formula hold for all them. We begin with an insertion of the identity, $1 = (1/V) \sum_s \int d^3 \mathbf{x} | \mathbf{x}, s \rangle \langle \mathbf{x}, s |$, where $V$ is the target volume,
\begin{align}
    \mathcal{T}_{i,f,\mathbf{k},\mathbf{k}'} = \frac{1}{V} \sum_{s}\int d^3\mathbf{x} \, e^{i ( \mathbf{q} - \mathbf{k}') \cdot \mathbf{x}} u_{f, \mathbf{k}', s}^*(\mathbf{x}) \langle \mathbf{x}, s | \hat{\mathcal{O}} | I \rangle \, ,
    \label{eq:T_pre_bloch_simplify}
\end{align}
where $\hat{\mathcal{O}}$ is an operator inside the $\mathcal{T}$'s in Eqs.~\eqref{eq:T_1} -~\eqref{eq:T_vxs} without the $e^{i \mathbf{q} \cdot \mathbf{x}}$ term. All of these $\hat{\mathcal{O}}$ are diagonal in momentum space, therefore it is simplest to insert a momentum space identity,
\begin{align}
    \langle \mathbf{x}, s | \hat{\mathcal{O}} | I \rangle & = V \sum_{s'} \int \frac{d^3\mathbf{p}}{(2\pi)^3} e^{i \mathbf{p} \cdot \mathbf{x}} \mathcal{O}_{ss'}(\mathbf{p}) \langle \mathbf{p}, s' | I \rangle \, .
    \label{eq:O_simplify}
\end{align}
One can further show that,
\begin{align}
    \langle \mathbf{p}, s | I \rangle = \frac{1}{V} \sum_{\mathbf{G}, s'} \widetilde{u}_{i, \mathbf{k}, s', \mathbf{G}} \delta^{3}(\mathbf{k} + \mathbf{G} - \mathbf{p}) \, ,
\end{align}
and therefore 
\begin{align}
    \langle \mathbf{x}, s | \hat{\mathcal{O}} | I \rangle & \equiv e^{i \mathbf{k} \cdot \mathbf{x}} \left[ \hat{\mathcal{O}} \cdot u \right]_{i, \mathbf{k}, s} = e^{i \mathbf{k} \cdot \mathbf{x}} \sum_{\mathbf{G}, s} \mathcal{O}_{s, s'}(\mathbf{k} + \mathbf{G}) \widetilde{u}_{i, \mathbf{k}, s', \mathbf{G}} e^{i \mathbf{G} \cdot \mathbf{x}} \, ,
    \label{eq:O_simplify_post}
\end{align}
where we have introduced some convenient shorthand notation $\left[ \right]$ for application of the operator in Fourier space, and $\mathcal{O}_{ss'}(\mathbf{p}) \equiv \langle \mathbf{p}, s | \hat{\mathcal{O}} | \mathbf{p}, s' \rangle $.\footnote{In words, the procedure to compute $\left[ \mathcal{O} \cdot u \right]$ is compute $u$, Fourier transform it to find $\widetilde{u}$, apply the $\mathcal{O}$ operator in momentum space, then Fourier transform back.}

Substituting Eq.~\eqref{eq:O_simplify_post} in to Eq.~\eqref{eq:T_pre_bloch_simplify}, and using the periodicity of $u$, then gives,
\begin{align}
    \mathcal{T}_{i,f,\mathbf{k}\mathbf{k}'} & = \frac{1}{V} \sum_s \int d^3\mathbf{x} e^{i (\mathbf{q} - \mathbf{k}' + \mathbf{k}) \cdot \mathbf{x}} u^*_{f, \mathbf{k}', s} \left[ \mathcal{O} \cdot u \right]_{i, \mathbf{k}, s} \\
                                            & = \sum_{\mathbf{G}} \delta_{\mathbf{q}, \mathbf{k}' - \mathbf{k} + \mathbf{G}} \frac{1}{\Omega} \sum_s \int_\text{UC} d^3\mathbf{x} \, e^{i \mathbf{G} \cdot \mathbf{x}}u^*_{f, \mathbf{k}', s} \left[ \mathcal{O} \cdot u \right]_{i, \mathbf{k}, s}
\end{align}
where UC indicates a sum of the unit cell. Note that, as a result of crystal lattice momentum conservation, $\mathcal{T}$ only has support at the kinematically allowed points, $\mathbf{q} = \mathbf{k}' - \mathbf{k} + \mathbf{G}$. Alternatively we can simplify this as,
\begin{align}
    \mathcal{T}_{i,f,\mathbf{k},\mathbf{k}'}(\mathbf{q} = \mathbf{k}' - \mathbf{k} + \mathbf{G}) & = \frac{1}{\Omega} \sum_s \int_\text{UC} d^3\mathbf{x} \, e^{i \mathbf{G} \cdot \mathbf{x}}u^*_{f, \mathbf{k}', s} \left[ \mathcal{O} \cdot u \right]_{i, \mathbf{k}, s} \, ,
\end{align}
which is the transition matrix element that gets computed for every pair of crystal electronic states.

\bibliographystyle{apsrev4-2}
\bibliography{bibliography}

\end{document}